\documentclass{article}
\usepackage[preprint]{log_2026}			

\usepackage{booktabs}						
\usepackage{multirow}						
\usepackage{amsfonts}						
\usepackage{graphicx}						
\usepackage{duckuments}						
\usepackage{amsmath}
\usepackage{algorithm}      
\usepackage{algpseudocode}  
\usepackage{xcolor}
\usepackage[subpreambles=true]{standalone}

\PassOptionsToPackage{numbers, compress}{natbib}
\usepackage[numbers,compress,sort]{natbib}	

\newcommand{\hp}{\texttt{\upshape HpStrat}}

\title[Not All Nodes Are Created Equal: Homophily-Aware Stratification for Stable GNN Evaluation]{Not All Nodes Are Created Equal: Homophily-Aware Stratification for Stable GNN Evaluation}

\author[Jami et al.]{%
    Naga Venkata Sai Jitin Jami$^{1,2,3,4}$\quad Thomas Altstidl$^{2,3,4}$\quad Sebastian Hoefler$^{2,3,4}$\\
  \normalsize\textbf{Jonas Leo Mueller$^{2,3,4}$ \quad Dario Zanca$^{2}$ \quad Björn Eskofier$^{2,3,4}$\quad Heike Leutheuser$^1$} \\[.4em]
  $^1$ Chair for Machine Learning in Medicine, Universität Bayreuth \\
  $^2$ Department Artificial Intelligence in Biomedical Engineering, FAU Erlangen-Nürnberg \\
  $^3$ Chair of AI-supported Therapy Decisions, Ludwig-Maximilians-Universität München \\
  $^4$ Munich Center for Machine Learning (MCML) \\
  \texttt{\{jitin.jami, heike.leutheuser\}@uni-bayreuth.de} \quad \texttt{eskofier@lmu.de} \\
  \texttt{\{thomas.r.altstidl,jonas.leo.mueller,sebastian.hoefler,dario.zanca\}@fau.de}
}

\begin{document}

\maketitle

\begin{abstract}
Graph neural networks are widely used for transductive node classification, with accuracy typically measured on randomly drawn train/validation/test splits. Reported accuracy has been shown to shift substantially across different random splits of the same dataset, making published comparisons between architectures unreliable. The classical remedy in non-graph settings is stratified $k$-fold cross-validation, which ensures each test fold reflects the full class distribution of the dataset. We argue that class stratification alone is insufficient for graphs: nodes are not isolated but connected, and folds that differ in their distribution of local neighbourhood homophily expose the model to systematically different relational conditions that directly affect message-passing behaviour. The resulting cross-fold variation reflects the homophily composition of each split, inflating reported variance beyond what model behaviour alone would produce. To address this, we propose \hp{}, a topology-aware stratification procedure that treats node homophily as the primary stratification axis, aligning folds with respect to local relational consistency alongside the class marginal that standard stratification already controls. Stratifying on homophily alone does not guarantee class balance, so \hp{} incorporates class label as a secondary axis, preserving class representativeness as a natural consequence of the procedure. We evaluate \hp{} on a broad benchmark suite comprising 15 node-classification datasets spanning the full homophily spectrum and 7 GNN architectures. \hp{} achieves a mean stability rank of 1.49 compared to 2.31 for random $k$-fold, achieving the lowest mean stability rank on 13 of 15 datasets while preserving class balance close to class-stratified splits and substantially better than random. We argue that homophily-aware split construction merits broader adoption for GNN evaluation. Code available at: \url{https://github.com/jitinjami/graph-strat.git}
\end{abstract}



\section{Introduction}
\label{sec:introduction}

Graph neural networks (GNNs) have become the standard tool for learning on relational data~\cite{gnnpaper}, with transductive node classification as a core problem setting: a model trains on a partially labelled graph and predicts labels for held-out nodes at inference time. This setting arises across citation and co-author networks~\cite{coradataset,yangplanetoid,pitfallspaper}, e-commerce co-purchase graphs~\cite{pitfallspaper,platanovchameleondataset}, Wikipedia and webpage networks~\cite{platanovchameleondataset,geomgcnpaper}, and social co-occurrence graphs~\cite{geomgcnpaper}. Over the years, successive architectures have been proposed and evaluated on these tasks~\cite{gcnpaper,gatpaper,sagepaper,appnppaper,mixhoppaper,h2gcnpaper,gprgnnpaper}. The reliability of this progress, however, has been called into question: standard benchmark datasets do not stress-test the capacity the field claims to measure~\cite{datasetpitfall}; modest hyper-parameter tuning of simpler models is sufficient to match or exceed more complex architectures~\cite{hyptunepitfall}; and reported accuracy shifts substantially under different random train/validation/test splits of the same data~\cite{pitfallspaper}. Evaluating across 200 randomly drawn splits has been proposed as a remedy for the latter~\cite{pitfallspaper}, but this does not scale to large graphs or models. More broadly, these critiques reflect a common pattern: evaluation protocols have been inherited from prior work and reused without scrutiny, with consequences that include, among others, undetected train-test leakage from duplicate nodes in widely used datasets~\cite{platanovchameleondataset}.

This documented split-induced performance instability~\cite{pitfallspaper}, however, has a remedy in classical Machine Learning: class stratified $k$-fold cross-validation~\cite{kuhnbook,kohavi1995study}, where the class distribution is preserved across train/val/test splits. Enforcing class balance across folds ensures that the model is evaluated in-distribution: each test set matches the datasets' class distribution rather than whatever composition a random split happens to produce. The mean accuracy across folds then estimates performance directly, and the standard deviation across folds indicates whether this estimate is stable or an artefact of the particular split drawn. Class stratification explicitly controls for this by construction, rather than relying on repeated random sampling to approximate it. For transductive node classification, however, class balance alone does not produce comparable test sets: nodes are not isolated entities, and their position within the graph also determines how much discriminative signal a GNN can extract from their neighbourhood. A split that is class-balanced but blind to this relational structure can still produce folds whose accuracies differ for reasons unrelated to the model. In particular, systematic differences in local neighbourhood structure between folds have been shown to distort a GNN's measured performance~\cite{homophily0}. Principled graph stratification therefore requires not only class balance but also a way to distribute the relational conditions of the problem uniformly across folds.

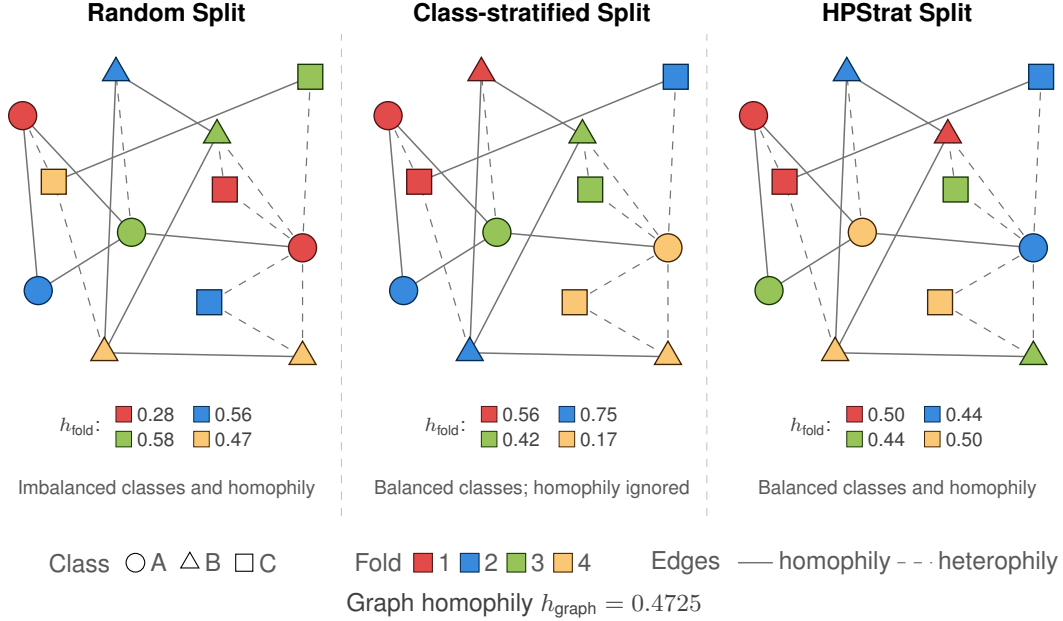
\begin{figure}[t]
\centering
\begin{tikzpicture}[x=1cm, y=1cm]

  \Panel{0}{}
  \begin{scope}[xshift=0cm]
    \node[clA, fA] at (n1)  {};
    \node[clA, fB] at (n2)  {};
    \node[clA, fC] at (n3)  {};
    \node[clB, fB] at (n4)  {};
    \node[clB, fD] at (n5)  {};
    \node[clC, fD] at (n6)  {};
    \node[clA, fA] at (n7)  {};
    \node[clB, fC] at (n8)  {};
    \node[clB, fD] at (n9)  {};
    \node[clC, fC] at (n10) {};
    \node[clC, fA] at (n11) {};
    \node[clC, fB] at (n12) {};
    \node[panellabel] at (2.15, 6.85) {Random Split};
    \FoldTable{2.15}{1.55}{0.28}{0.56}{0.58}{0.47}
    \node[caption]    at (2.15, 0.75) {Imbalanced classes and homophily};
  \end{scope}

  \Panel{4.7}{}
  \begin{scope}[xshift=4.7cm]
    \node[clA, fA] at (n1)  {};
    \node[clA, fB] at (n2)  {};
    \node[clA, fC] at (n3)  {};
    \node[clB, fA] at (n4)  {};
    \node[clB, fB] at (n5)  {};
    \node[clC, fA] at (n6)  {};
    \node[clA, fD] at (n7)  {};
    \node[clB, fC] at (n8)  {};
    \node[clB, fD] at (n9)  {};
    \node[clC, fB] at (n10) {};
    \node[clC, fC] at (n11) {};
    \node[clC, fD] at (n12) {};
    \node[panellabel] at (2.15, 6.85) {Class-stratified Split};
    \FoldTable{2.15}{1.55}{0.56}{0.75}{0.42}{0.17}
    \node[caption]    at (2.15, 0.75) {Balanced classes; homophily ignored};
  \end{scope}

  \Panel{9.4}{}
  \begin{scope}[xshift=9.4cm]
    \node[clA, fA] at (n1)  {};
    \node[clA, fC] at (n2)  {};
    \node[clA, fD] at (n3)  {};
    \node[clB, fB] at (n4)  {};
    \node[clB, fD] at (n5)  {};
    \node[clC, fA] at (n6)  {};
    \node[clA, fB] at (n7)  {};
    \node[clB, fA] at (n8)  {};
    \node[clB, fC] at (n9)  {};
    \node[clC, fB] at (n10) {};
    \node[clC, fC] at (n11) {};
    \node[clC, fD] at (n12) {};
    \node[panellabel] at (2.15, 6.85) {HPStrat Split};
    \FoldTable{2.15}{1.55}{0.50}{0.44}{0.44}{0.50}
    \node[caption]    at (2.15, 0.75) {Balanced classes and homophily};
  \end{scope}

  \draw[black!25, dashed, line width=0.3pt] (4.4, 0.4) -- (4.4, 6.6);
  \draw[black!25, dashed, line width=0.3pt] (9.1, 0.4) -- (9.1, 6.6);

  \begin{scope}[yshift=-0.2cm]
    \node[legendlabel, anchor=east] at (1.55, 0) {Class};
    \node[clALeg, fill=white] at (1.75, 0) {};
    \node[legendlabel, anchor=west, inner sep=1pt] at (1.9, 0) {A};
    \node[clBLeg, fill=white] at (2.45, 0) {};
    \node[legendlabel, anchor=west, inner sep=1pt] at (2.6, 0) {B};
    \node[clCLeg, fill=white] at (3.15, 0) {};
    \node[legendlabel, anchor=west, inner sep=1pt] at (3.3, 0) {C};

    \node[legendlabel, anchor=east] at (5.3, 0) {Fold};
    \node[draw=strokeA, fill=foldA, line width=0.4pt, minimum size=2.4mm, inner sep=0pt] at (5.45, 0) {};
    \node[legendlabel, anchor=west, inner sep=1pt] at (5.6, 0) {1};
    \node[draw=strokeB, fill=foldB, line width=0.4pt, minimum size=2.4mm, inner sep=0pt] at (6.05, 0) {};
    \node[legendlabel, anchor=west, inner sep=1pt] at (6.2, 0) {2};
    \node[draw=strokeC, fill=foldC, line width=0.4pt, minimum size=2.4mm, inner sep=0pt] at (6.65, 0) {};
    \node[legendlabel, anchor=west, inner sep=1pt] at (6.8, 0) {3};
    \node[draw=strokeD, fill=foldD, line width=0.4pt, minimum size=2.4mm, inner sep=0pt] at (7.25, 0) {};
    \node[legendlabel, anchor=west, inner sep=1pt] at (7.4, 0) {4};

    \node[legendlabel, anchor=east] at (9.4, 0) {Edges};
    \draw[homoedge] (9.50, 0) -- (9.95, 0);
    \node[legendlabel, anchor=west, inner sep=2pt] at (9.95, 0) {homophily};
    \draw[heteroedge] (11.55, 0) -- (12.00, 0);
    \node[legendlabel, anchor=west, inner sep=2pt] at (12.00, 0) {heterophily};

    \node[font=\sffamily\footnotesize, text=black!75, anchor=center] at (6.75, -0.55)
      {Graph homophily $h_{\text{graph}} = 0.4725$};
  \end{scope}

\end{tikzpicture}
\caption{A toy graph with three node classes and graph-level homophily $h_{graph}=0.4725$, partitioned into $k=4$ folds by three strategies, with per-fold mean homophily shown. Random $k$-fold (left) yields imbalanced classes and high fold-homophily variance (CHD, Eq.~\ref{eq:chd}, $=0.0975$). Class-stratified $k$-fold (centre) recovers class balance but leaves homophily unconstrained (CHD $=0.180$). \hp{} (right) controls both, preserving class balance while CHD falls to $0.030$.}
\label{fig:overview}
\end{figure}

The relational context a node inhabits is captured by its local neighbourhood structure, and understanding how GNNs interact with this structure is key to identifying the right stratification axis. GNNs belong to a broader family of models known as Message Passing Neural Networks (MPNNs)~\cite{gnnismpnn,adjustedhomophily}, which update each node's representation by aggregating information from its local neighbourhood at every layer. Classical instantiations such as GCN~\cite{gcnpaper}, GAT~\cite{gatpaper}, and GraphSAGE~\cite{sagepaper} realise this through simple averaging or concatenation of neighbour features combined with graph topology. The topology of the graph therefore acts as a source of inductive bias~\cite{inductivebiases}: MPNNs are designed to implicitly work on the assumption that connected nodes tend to be similar to one another, an assumption known as \textit{homophily}~\cite{birds}. Standard GNN designs rely on it by construction, a dependence made explicit in the heterophily literature that followed~\cite{geomgcnpaper,h2gcnpaper}, which identified homophily dependence as a central constraint of standard GNN architectures and motivated new architectures built specifically to relax it.

The relationship between homophily and GNN performance, however, is not straightforward. Early work attributed poor GNN performance on heterophilous graphs to the misalignment between neighbourhood aggregation and label structure, motivating architectures explicitly designed to handle heterophily~\cite{geomgcnpaper,h2gcnpaper,coin}. Subsequent work showed that homophily is not a necessary condition for strong performance, documenting a non-monotone relationship between global homophily and accuracy that depends on neighbourhood label distributions rather than homophily level alone~\cite{homophily1,homophily3,homophily5}. Crucially, performance discrepancies arise not only from global homophily but from local deviations: when a node's local homophily differs from the global homophily of the training graph, GNN predictions are distorted in ways that are theoretically predictable~\cite{homophily0}. This sensitivity means that test sets with systematically different homophily compositions produce performance differences that are an artefact of split construction rather than a signal about the model. In Appendix~\ref{app:var_decomp}, we confirm this empirically across multiple datasets and model combinations: holding the model definition fixed, test accuracy varies significantly ($p<0.05$, $R^2 \in [0.021, 0.289]$) across randomly drawn splits with different test-set homophily. Node homophily, \(h_v\), is therefore an appropriate candidate for the stratification axis: it captures the relational context that governs GNN behaviour in a single scalar derived almost entirely from labels, with only minimal dependence on local neighbourhood structure. Crucially, it has an advantage that other structural properties lack: degree, centrality, and spectral position all expose topology that GNNs learn from directly, making them unsuitable as stratification keys. Stratifying on $h_v$ therefore controls the axis along which GNN performance is sensitive without leaking the structural information the model itself takes as input.

Motivated by these properties, we propose \hp{}, a stratification procedure that constructs $k$-fold splits by controlling for node homophily and class label jointly, extending the stratification principle from the class marginal to the relational structure of the graph. The resulting procedure preserves the homophily distribution in each fold while class representativeness follows as a natural consequence. It requires no changes to model training or architecture, serves as a drop-in replacement for random $k$-fold, adds no additional training runs, and produces a deterministic partition for a fixed graph and seed. To evaluate \hp{}, we construct an empirical study spanning 15 transductive node classification datasets that cover the full homophily spectrum, from strongly homophilous citation networks to strongly heterophilous webpage and co-occurrence graphs. We evaluate 7 GNN architectures: GCN~\cite{gcnpaper}, GAT~\cite{gatpaper}, GraphSAGE~\cite{sagepaper}, APPNP~\cite{appnppaper}, MixHop~\cite{mixhoppaper}, H2GCN~\cite{h2gcnpaper}, and GPR-GNN~\cite{gprgnnpaper}, ranging from homophily-assuming to heterophily-robust designs. We compare against random $k$-fold and class-stratified $k$-fold using metrics designed to assess fold quality on size, class distribution, and average homophily, as well as stability of test accuracy across folds. \hp{} achieves lower cross-fold standard deviation than both baselines across the full homophily spectrum and all evaluated architectures. Class-stratified $k$-fold improves over random splitting on a subset of datasets but not consistently, suggesting that controlling the class marginal alone is insufficient to stabilise GNN evaluation. Controlling the relational structure is the more consequential intervention. 
\section{Related Work}
\label{sec:related_work}

\paragraph{Stratified sampling and iterative stratification.}
Stratified $k$-fold cross-validation~\cite{kohavi1995study,kuhnbook} is the standard answer to evaluation instability in classification: by preserving class proportions across folds, it ensures each test set is representative of the overall label distribution. Extending stratification beyond single-label settings has been an active research direction. A greedy iterative procedure that assigns samples to folds by satisfying demand for the rarest label first was proposed for multi-label data~\cite{sechidis2011stratification}, and later extended to account for second-order label co-occurrence relationships~\cite{szymanski2017network}. A pre-sorting variant that clusters samples with similar label sets before fold assignment achieves improved label homogeneity~\cite{pmbsrs}. For extreme multi-label datasets where exact balance is computationally intractable, a scalable approximate stratification procedure has been proposed~\cite{ssstrat}. Dedicated fold quality measures for multi-label settings alongside a direct optimisation algorithm, Optisplit, have also been developed~\cite{optisplit}. A genetic-algorithm formulation, EvoSplit, jointly optimises label and label-pair distribution across folds~\cite{evosplit}. Most recently, stratification has been extended to image segmentation, where each sample carries a dense pixel-level label map: a pixel-aware greedy procedure and a Wasserstein-distance-minimising evolutionary algorithm have been shown to produce more representative splits than random assignment~\cite{jami2025}. None of these methods address the graph setting, where the relevant stratification axis is not a label or label combination but a relational property of nodes derived from graph structure. The mismatch is structural rather than incidental. The evolutionary and Wasserstein-based procedures~\cite{evosplit,jami2025} are designed for complex, multi-dimensional target distributions with several interacting labels, whereas here there is a single scalar distribution to match. The multi-label and second-order stratifiers~\cite{sechidis2011stratification,szymanski2017network,pmbsrs,ssstrat,optisplit} are specfically designed for discrete label co-occurrence structure, whereas node homophily is a continuous variable. \hp{} addresses this gap by controlling for discretised node homophily jointly with class label, ensuring folds are balanced on the relational property that governs GNN performance rather than on class distribution alone.

\paragraph{Benchmark datasets and standard splits.}
The Planetoid datasets (Cora, CiteSeer, and PubMed~\cite{coradataset}) are among the most widely used benchmarks in transductive node classification. The original splits were designed for semi-supervised learning, using a small fixed training set of 20 nodes per class, 500 nodes for validation, and 1000 nodes for testing~\cite{yangplanetoid}, and were adopted widely~\cite{gcnpaper,gatpaper,monetpaper,appnppaper}. Randomly drawn splits under the same semi-supervised regime of 20 nodes per class for training and 30 nodes per class for validation were proposed~\cite{pitfallspaper} and later adopted~\cite{gprgnnpaper}. As the field moved toward supervised evaluation, larger training sets became more common: MixHop~\cite{mixhoppaper} used 100 nodes per class for training and 500 for validation with the remainder as test nodes. A percentage-based 60/20/20 protocol applied to the full dataset, giving rise to a substantially larger training set, was also proposed~\cite{geomgcnpaper}, though it was later noted to be effectively 50/30/20~\cite{h2gcnpaper}. The WebKB datasets (Cornell, Wisconsin, Texas), Actor, and the Wikipedia network datasets (Chameleon and Squirrel) were introduced in the same publication alongside this protocol~\cite{geomgcnpaper}, and the splits were adopted in subsequent work~\cite{h2gcnpaper,homophily1}. Chameleon and Squirrel were subsequently found to contain duplicate nodes causing train-test leakage; revised versions with corrected splits were proposed as replacements~\cite{platanovchameleondataset}. The Amazon (Photo, Computers) and CoAuthor (CS, Physics) datasets were introduced in the same publication that proposed the random semi-supervised splits~\cite{pitfallspaper}, under the same 20 nodes per class training and 30 nodes per class validation protocol. A balanced 50/25/25 protocol covering all of the above datasets was proposed~\cite{homophily0} and adopted~\cite{axdependence,trihom}. However, no consistent split-construction practice has emerged across the literature: some protocols were released as fixed splits~\cite{yangplanetoid,geomgcnpaper,pyg}, ensuring reproducibility, while others were described procedurally with no accompanying implementation, making exact replication difficult. None of these efforts propose a general split-construction procedure applicable to arbitrary transductive node classification graphs. \hp{} is a dataset-agnostic procedure that produces representative $k$-fold splits for any labelled graph, without modifying the underlying dataset or requiring domain-specific split design.

\section{Method}
\label{sec:method}

\subsection{Problem setup and notation}
\label{sec:method:setup}

Consider a transductive node classification task on an undirected, unweighted graph \(\mathcal{G} = (\mathcal{V}, \mathcal{E})\) with node set \(\mathcal{V}\) of \(n\) nodes and edge set \(\mathcal{E}\). Each node \(v\) has a class label \(y_v \in \{1, \ldots, C\}\); the full label vector is \(\mathbf{Y} \in \{1, \ldots, C\}^n\). For a node \(v \in \mathcal{V}\), let \(\mathcal{N}(v) = \{u \in \mathcal{V} : (u, v) \in \mathcal{E}\}\) denote its immediate neighbours. Node homophily, the fraction of a node's neighbours that share its label, of node \(v\) is:
\begin{equation}
h_v \;=\; \frac{|\{u \in \mathcal{N}(v) : y_u = y_v\}|}{|\mathcal{N}(v)|} \;\in\; [0,1], \qquad |\mathcal{N}(v)| > 0,
\label{eq:node_homophily}
\end{equation}
Computing \(h_v\) requires the label of node \(v\) and its neighbours. This is consistent with class-wise stratification in the transductive setting, where labels are hidden during model training but not during train/val/test split. The objective is to partition the nodeset \(\mathcal{V}\) into \(k\) equally sized disjoint subsets \(\mathbf{S} = \{S_1, \ldots, S_k\}\). For \(k\) evaluation rounds in a \(k\)-fold cross-validation setup: one fold serves as the test set, one adjacent fold serves as the validation set, and the remaining \(k-2\) folds form the training set.

\subsection{\hp{}: Homophily-Priority Stratification}
\label{sec:method:hp_priority}
Let \(\mathbf{h} = \{h_v : v \in \mathcal{V}\}\) denote the homophily values of all nodes and \(\mathcal{P}_h\) denote the empirical distribution of \(\mathbf{h}\). These are discretised into \(B\) equal-width bins. A node \(v\) belonging to bin \(b\) is denoted \(v_b\); the set of all nodes in bin \(b\) is \(\mathcal{V}_b\). A node in bin \(b\) with class label \(c\) is denoted \(v_b^c\), with corresponding stratum \(\mathcal{V}_{b,c} = \{v \in \mathcal{V}_b : y_v = c\}\) for \(c \in \{1, \ldots, C\}\).

As detailed in Algorithm~\ref{alg:hp_priority} in Appendix~\ref{app:algorithm}, \hp{} iterates through each bin \(\mathcal{V}_b\) in order. Within each bin, it iterates through the classes present and processes each stratum \(\mathcal{V}_{b,c}\) in turn. For each stratum, the fold with the fewest currently assigned nodes is identified as the starting fold; when multiple folds are tied, the lowest-indexed fold is selected. The first node in the stratum is assigned to the starting fold, the second to the next fold, and so on in sequence, wrapping back to the first fold after the last, until all nodes in the stratum are exhausted. Choosing the fold with the fewest nodes at the beginning of each stratum ensures that any imbalance accumulated from previous strata is corrected rather than propagated. Once all strata in a bin are processed, each fold holds a proportional representation of the class distribution within that bin. After all bins are processed, the empirical homophily distribution \(\mathcal{P}_h\) is preserved across folds by construction, and class balance is achieved as a cumulative consequence of the per-bin assignments.

\begin{figure}[t]
\centering
\includegraphics[width=\textwidth]{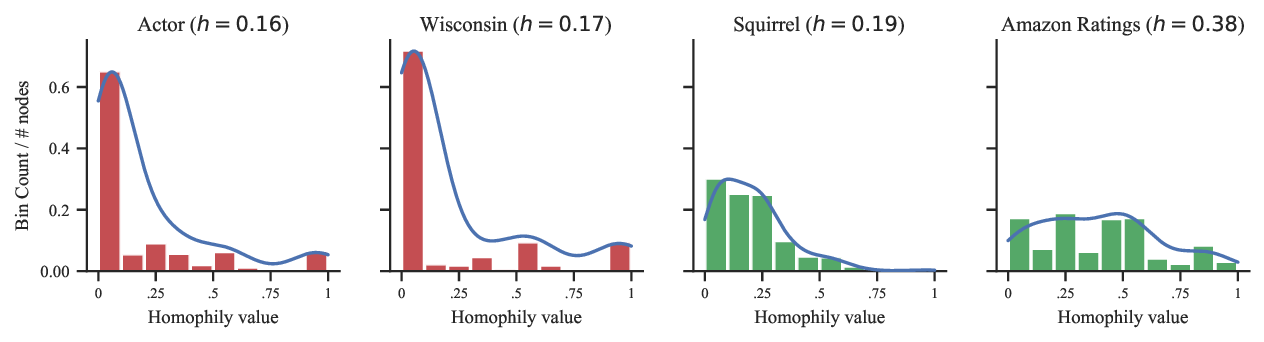}
\caption{Node homophily distributions for four datasets with comparable mean homophily but different distribution shapes. Wisconsin ($\bar{h}=0.17$) and Actor ($\bar{h}=0.16$) concentrate most nodes in the lowest bins; Amazon-Ratings ($\bar{h}=0.38$) and Squirrel ($\bar{h}=0.19$) spread nodes across multiple bins. This bin skew, not mean homophily, explains why \hp{} degrades on Wisconsin and Actor but not the other two.}
\label{fig:homophily_distributions}
\end{figure}

\hp{} applies where meaningful homophily variation exists across nodes. When the homophily distribution is degenerate (all nodes fall into a single bin), every stratum \(\mathcal{V}_{b,c}\) reduces to a class stratum \(\mathcal{V}_c\) and the outer loop over bins executes exactly once. The inner loop then iterates over classes and distributes nodes via round-robin in precisely the same way as class-stratified $k$-fold, making the two procedures identical in this case. The same behaviour emerges to a lesser degree under skewed homophily distributions as illustrated in Figure~\ref{fig:homophily_distributions}: Amazon-Ratings and Squirrel have support across multiple bins, allowing \hp{} to distribute nodes representatively, while Wisconsin and Actor concentrate most nodes in the lowest bins, where the procedure converges to class-stratified $k$-fold.

\subsection{Fold-quality metrics}
\label{sec:method:fold_metrics}

To assess whether each splitting procedure produces folds that are representative of the full graph and well-formed as evaluation partitions, we propose three metrics computed directly from the fold structure. These metrics quantify: (1) how closely each fold's size matches the target proportion of $n/k$ of the nodes; (2) how well each fold's class distribution matches that of the full graph; and (3) how closely the average homophily of each test fold matches that of the whole dataset.

\textbf{Sample Distribution Deviation (SDD)} measures how uniformly nodes are distributed across folds. In an ideal partition, each fold receives exactly $n/k$ nodes; in practice, this is not always achievable when $n$ is not divisible by $k$ or when stratification constraints force uneven assignments. SDD quantifies the mean absolute deviation of each fold's size from this target:
\begin{equation}
\mathrm{SDD} \;=\; \frac{1}{k} \sum_{i=1}^k \left| |S_i| - \frac{n}{k} \right|.
\label{eq:sdd}
\end{equation}
A low SDD indicates that folds are of comparable size, ensuring that each evaluation round uses a consistent number of nodes in test data.

\textbf{Class Distribution Deviation (CDD)} measures how much the class representation within each fold deviates from the expected count of $|\mathcal{V}_c| / k$ nodes per class. In an ideal partition, each fold receives exactly this many nodes from each class; in practice, indivisibility and stratification constraints introduce deviations from this target. CDD quantifies the mean absolute deviation from the expected class count, averaged over folds and classes:
\begin{equation}
\mathrm{CDD} \;=\; \frac{1}{C} \sum_{c=1}^C \frac{1}{k} \sum_{i=1}^k \left| |S_i \cap \mathcal{V}_c| \;-\; \frac{|\mathcal{V}_c|}{k} \right|.
\label{eq:cdd}
\end{equation}
A low CDD indicates that each fold presents a class distribution close to that of the full dataset, ensuring that each evaluation round is equally representative of the classification problem and that variance in measured accuracy is not driven by class imbalance in the test set.

\textbf{Cross-Fold Homophily Deviation (CHD)} measures how closely the mean homophily of each fold matches that of the full graph. In an ideal partition, every fold has the same mean homophily as the dataset as a whole. Split strategies that do not control for homophily risk producing folds whose neighbourhood structure deviates systematically from the global distribution, making some test sets harder or easier for reasons unrelated to model capability. CHD quantifies this as the mean absolute deviation of per-fold mean homophily from the global mean:
\begin{equation}
\mathrm{CHD} \;=\; \frac{1}{k} \sum_{i=1}^{k} \left| \frac{1}{|S_i|} \sum_{v \in S_i} h_v \;-\; \frac{1}{|\mathcal{V}|} \sum_{v \in \mathcal{V}} h_v \right|.
\label{eq:chd}
\end{equation}
A low CHD indicates that each fold is representative of the global homophily regime, ensuring that no fold presents neighbourhood conditions that systematically diverge from the dataset as a whole.

By design, random $k$-fold is expected to perform best on SDD and class-stratified $k$-fold on CDD, as each controls the relevant property explicitly. Neither baseline controls for homophily, so CHD is the metric where the effect of \hp{} is most directly observable. CHD is a sample mean, and sample means become more stable as graph size grows. This means the gap between random $k$-fold and \hp{} on CHD specifically is expected to narrow on larger graphs. However, \hp{} controls the full per-bin fold composition, not just the global mean, so its advantage in distributional representativeness should persist even as the mean gap narrows. SDD and CDD additionally let us verify whether \hp{} degrades fold size uniformity or class representativeness relative to the baselines, the latter being of particular interest since class balance is only a secondary axis of \hp{}, controlled implicitly through per-stratum round-robin assignment.
\section{Experimental Setup}
\label{sec:experimental_setup}

\subsection{Datasets}
\label{sec:setup:datasets}

A meaningful evaluation of split methods requires datasets that stress-test the homophily axis. We perform our analysis on 15 transductive node classification datasets spanning the full homophily spectrum, from citation networks with $h > 0.8$ to heterophilous web graphs with $h < 0.1$. All datasets are sourced from the PyTorch Geometric library~\cite{pyg}. The exception is Chameleon and Squirrel, for which we use the corrected versions from source~\cite{platanovchameleondataset}\footnote{\url{https://github.com/yandex-research/heterophilous-graphs/}}, which removed duplicate nodes present in the original graphs that caused train-test leakage. Node count (\(|\mathcal{V}|\)), edge count (\(|\mathcal{E}|\)), number of classes (\(C\)), and mean node homophily (\(h\)) for each dataset are summarised in Table~\ref{tab:datasets}.

\begin{table}[t]
\centering
\caption{Properties of the 15 benchmark datasets, ordered from highest to lowest mean node homophily $\bar{h} = \frac{1}{|\mathcal{V}|}\sum_v h_v$. Ratings = Amazon-Ratings; Roman = Roman-Empire.}
\label{tab:datasets}
\resizebox{\textwidth}{!}{%
\begin{tabular}{lccccccccccccccc}
\toprule
 & Physics & Photo & CS & Cora & PubMed & Computers & CiteSeer & Ratings & Chameleon & Squirrel & Wisconsin & Actor & Cornell & Texas & Roman \\
\midrule
\(|\mathcal{V}|\)  & 34{,}493 & 7{,}650 & 18{,}333 & 2{,}708 & 19{,}717 & 13{,}752 & 3{,}327 & 24{,}492 & 890 & 2{,}223 & 251 & 7{,}600 & 183 & 183 & 22{,}662 \\
\(|\mathcal{E}|\)  & 495{,}924 & 238{,}162 & 163{,}788 & 10{,}556 & 88{,}648 & 491{,}722 & 9{,}104 & 186{,}100 & 17{,}708 & 93{,}996 & 515 & 30{,}019 & 298 & 325 & 65{,}854 \\
\(C\)              & 5 & 8 & 15 & 7 & 3 & 10 & 6 & 5 & 5 & 5 & 5 & 5 & 5 & 5 & 18 \\
\(\bar{h}\)              & 0.92 & 0.84 & 0.83 & 0.83 & 0.79 & 0.79 & 0.71 & 0.38 & 0.24 & 0.19 & 0.17 & 0.16 & 0.11 & 0.07 & 0.05 \\
\bottomrule
\end{tabular}}
\end{table}

\subsection{Models and hyperparameters}
\label{sec:setup:models}

We evaluate on two groups of architectures. The first group consists of widely adopted baselines from the literature: GCN~\cite{gcnpaper}, GAT~\cite{gatpaper}, GraphSAGE~\cite{sagepaper}, and MixHop~\cite{mixhoppaper}. The second group consists of architectures that explicitly address the relationship between homophily and GNN performance: APPNP~\cite{appnppaper}, H2GCN~\cite{h2gcnpaper}, and GPR-GNN~\cite{gprgnnpaper}. Pytorch Geometric~\cite{pyg} reference implementations are used where available; H2GCN and GPR-GNN implementations are taken from literature~\cite{linkx}\footnote{\url{https://github.com/cuai/non-homophily-large-scale}} and integrated into a shared training loop.

All models are trained with \texttt{2} layers, \texttt{64} hidden units, dropout \texttt{0.5}, the Adam optimiser with learning rate \texttt{0.01} and weight decay \texttt{5e-4}, for a maximum of \texttt{1000} epochs with cross entropy loss. Early stopping is set at patience of \texttt{100} epochs on validation loss. All experiments are conducted on a single Nvidia A100 graphics card with 40GB of VRAM. Model-specific parameters are fixed as follows: GAT uses \(8 \times 8\) attention heads; GraphSAGE uses \texttt{mean} aggregation; APPNP uses \(K=10\) propagation steps and teleport \(\alpha=0.1\); MixHop uses adjacency powers \(\{0,1,2\}\); H2GCN uses the H2GCN-2 variant; GPR-GNN uses \(K=10\) propagation steps and PPR initialisation of \(\alpha=0.1\). We set \(B=10\) bins for \hp{}; Appendix~\ref{app:bin_ablation} reports a sensitivity analysis on \(B\). This shared configuration, taken directly from the literature~\cite{h2gcnpaper}, is held fixed across every dataset and split method rather than tuned separately for each. We choose not to tune per dataset or per split method, since doing so would introduce circularity: the validation set used for tuning is itself constructed under one of the three strategies being compared, making the tuned model no longer a neutral basis for judging that strategy. Holding these configurations fixed across all datasets and split methods isolates the effect of the splitting protocol as the only varying factor.

\subsection{Splits and evaluation protocol}
\label{sec:setup:splits}

Default masks shipped with each dataset are not used. Across the 15 datasets, split protocols are inconsistent: Cora, CiteSeer, and PubMed come with a single fixed Planetoid mask~\cite{yangplanetoid}; Cornell, Wisconsin, Texas, Actor, Chameleon, and Squirrel are reported with 10 fixed splits~\cite{geomgcnpaper}; Amazon-Ratings and Roman are reported with fixed splits~\cite{platanovchameleondataset}; and Amazon (Photo, Computers) and CoAuthor (CS, Physics) were introduced with no fixed masks, only a split percentage described in prose~\cite{pitfallspaper}. More fundamentally, a single fixed mask is incompatible with cross-fold variance estimation regardless of how it is constructed. We construct \(k=4\) folds with a \(50\%/25\%/25\%\) train/validation/test split, the closest multiple of \(1/k\) to the \(48\%/32\%/20\%\) split used in prior work~\cite{h2gcnpaper}. We additionally repeat all experiments at \(k=5\) and at \(k=4\) averaged over 6 random seeds (24 runs total), and report in Appendix~\ref{app:acc_per_dataset}. Under this protocol, partitions are constructed independently under each of the three methods compared: random \(k\)-fold, class-stratified \(k\)-fold, and \hp{}, using \texttt{scikit-learn} for the first two and a \texttt{numpy}-based implementation on top of \texttt{scikit-learn} for \hp{}. For each fold, the model is trained on the training set, early stopping is applied based on validation loss, and the test set is evaluated exactly once at the end of training. The standard deviation of test accuracy across the \(k\) folds is then recorded as the primary evaluation metric for each combination of dataset, model, and split method.
\section{Results and Discussion}
\label{sec:results}

\subsection{Ranking procedure}
\label{sec:results:ranking}

To aggregate results across models and datasets, the three split methods are ranked by cross-fold standard deviation for each combination of dataset and model, with rank 1 assigned to the lowest variance. These ranks are then averaged along two axes: across all 7 models for each dataset, producing a per-dataset mean rank, and across all 15 datasets for each model, producing a per-model mean rank. Both are then averaged globally to produce a single aggregate rank per method. A lower rank along any axis indicates that a method more consistently produces stable evaluation in that setting. The same ranking procedure is applied independently to each of the three model-free fold-quality metrics (SDD, CDD, and CHD) defined in Section~\ref{sec:method:fold_metrics}, where rankings are computed on mean values across 10 seeds and rank 1 is assigned to the method closest to the ideal value.


\subsection{Fold-quality metrics}
\label{sec:results:fold_quality}

\begin{table}[t]
\centering
\caption{Aggregate fold-quality rankings across all 15 datasets based on mean values across 10 seeds (lower is better). \hp{} ties for best on SDD, confirming round-robin assignment does not sacrifice fold size uniformity. On CHD, class-stratified $k$-fold offers no meaningful improvement over random $k$-fold, confirming that class stratification does not account for homophily.}
\label{tab:fold_quality}
\begin{tabular}{lccc}
\toprule
Strategy & SDD & CDD & CHD \\
\midrule
Random           & \textbf{1.00} & 3.00          & 2.53 \\
Class-stratified & \textbf{1.00} & \textbf{1.00} & \textit{2.47} \\
\hp{}            & \textbf{1.00} & \textit{2.00} & \textbf{1.00} \\
\bottomrule
\end{tabular}
\end{table}

Per-dataset fold-quality metric values, reported as mean and standard deviation across 10 seeds, are provided in Appendix~\ref{app:foldmetrics_per_dataset}. Table~\ref{tab:fold_quality} reports the aggregate rankings across all 15 datasets. All three methods produce equally sized folds, as reflected by a three-way tie on SDD; the round-robin assignment used by both stratified methods does not come at a cost to fold size uniformity. Class-stratified $k$-fold achieves the best CDD on every dataset, as expected given that class balance is its explicit objective. On CHD, \hp{} ranks first on every dataset with a mean rank of 1.00, confirming that the homophily-priority assignment successfully preserves the empirical homophily distribution across folds. Notably, \hp{} ranks second on CDD despite class balance being only a secondary objective: per-dataset CDD values remain close to those of class-stratified $k$-fold and substantially below random $k$-fold, suggesting that the per-stratum round-robin assignment recovers class representativeness as a natural consequence of the procedure.

On graph data, two nodes carrying the same label can sit in radically different local neighbourhoods, contributing differently to GNN evaluation regardless of their shared label. The CHD results quantify this directly: class-stratified $k$-fold offers no meaningful improvement over random $k$-fold on this metric. The homophily distribution is the axis along which evaluation variance is most consequential for graph data, and class stratification does not control it.

\subsection{Cross-fold stability of test accuracy}
\label{sec:results:accuracy}

Table~\ref{tab:permodel_agg} reports the aggregate ranking per model, averaged across all 15 datasets. \hp{} achieves the lowest overall mean rank (1.49), ahead of class-stratified \(k\)-fold (2.20) and random \(k\)-fold (2.31). \hp{} ranks first for every individual model, with per-model mean ranks ranging from 1.27 (H2GCN) to 1.80 (GCN). The consistency holds across architecture families: \hp{} produces more stable evaluation not only for heterophily-robust models such as H2GCN and GPR-GNN, but also for standard aggregation-based models such as GCN and GAT that do not explicitly account for homophily. On average across models, class-stratified \(k\)-fold and random \(k\)-fold remain tightly clustered, suggesting that class stratification alone is not a sufficient intervention for graph evaluation stability.

Table~\ref{tab:petdataset_agg} reports per-dataset rankings averaged across all 7 models. \hp{} achieves the lowest mean rank on 13 of 15 datasets, spanning the full homophily spectrum from strongly homophilous graphs such as Physics (\(h=0.92\)) and CS (\(h=0.83\)) to strongly heterophilous graphs such as Roman-Empire (\(h=0.05\)) and Texas (\(h=0.07\)). On Actor, class-stratified $k$-fold ranks lowest. On Wisconsin, random and class-stratified $k$-fold tie for the lowest rank, with \hp{} ranking third. Full per-dataset accuracy and standard deviation values are provided in Appendix~\ref{app:acc_per_dataset}.

\begin{table}[t]
\centering
\caption{Per-model aggregate rankings by cross-fold standard deviation of test accuracy, averaged across all 15 datasets (rank 1 = lowest variance). \hp{} ranks first on every architecture, and the gap over both baselines is consistent across homophily-assuming (GCN, GAT) and heterophily-robust (H2GCN, GPR-GNN) designs.}
\label{tab:permodel_agg}
\resizebox{\textwidth}{!}{%
\begin{tabular}{lccccccc|c}
\toprule
Strategy & GCN & GAT & GraphSAGE & APPNP & MixHop & H2GCN & GPR-GNN & Mean Rank \\
\midrule
Random           & 2.20          & 2.47          & \textit{2.27} & \textit{2.20} & \textit{2.00} & \textit{2.33} & 2.73          & 2.31 \\
Class-stratified & \textit{2.00} & \textit{2.00} & 2.40          & 2.40          & 2.27          & 2.40          & \textit{1.93} & \textit{2.20} \\
\hp{}            & \textbf{1.80} & \textbf{1.53} & \textbf{1.33} & \textbf{1.40} & \textbf{1.73} & \textbf{1.27} & \textbf{1.33} & \textbf{1.49} \\
\bottomrule
\end{tabular}}
\end{table}

\begin{table}[t]
\centering
\caption{Per-dataset strategy rankings by cross-fold standard deviation of test accuracy, averaged across all 7 models (rank 1 = lowest variance). \hp{} achieves the lowest mean rank across all datasets. The two exceptions, Wisconsin and Actor, are the datasets with strongly skewed homophily distributions concentrated in the lowest bins. Datasets ordered high to low homophily. Ratings = Amazon-Ratings; Roman = Roman-Empire.}
\label{tab:petdataset_agg}
\resizebox{\textwidth}{!}{%
\begin{tabular}{lrrrrrrrrrrrrrrr|r}
\toprule
Strategy & Physics & Photo & CS & Cora & PubMed & Computers & CiteSeer & Ratings & Chameleon & Squirrel & Wisconsin & Actor & Cornell & Texas & Roman & Mean Rank \\
\midrule
Random           & 2.86          & \textit{1.86} & \textit{2.14} & 2.86          & 2.43          & 2.29          & 2.71          & \textit{2.14} & \textit{2.14} & 2.71          & \textbf{1.86} & 2.14          & 2.57          & \textit{1.86} & \textit{1.86} & 2.31 \\
Class-stratified & \textit{1.71} & 3.00          & 2.86          & \textit{2.00} & \textit{1.86} & \textit{2.14} & \textit{1.71} & 2.29          & 2.57          & \textit{1.86} & \textbf{1.86} & \textbf{1.86} & \textit{2.00} & 2.71          & 2.57          & \textit{2.20} \\
\hp{}            & \textbf{1.43} & \textbf{1.14} & \textbf{1.00} & \textbf{1.14} & \textbf{1.43} & \textbf{1.57} & \textbf{1.57} & \textbf{1.57} & \textbf{1.29} & \textbf{1.43} & 2.29          & \textit{2.00} & \textbf{1.43} & \textbf{1.43} & \textbf{1.57} & \textbf{1.49} \\
\bottomrule
\end{tabular}}
\end{table}

Across a majority of datasets, the method that achieves the lowest CHD also achieves the lowest cross-fold standard deviation of test accuracy, consistent with the hypothesis that homophily distribution mismatch is a primary driver of evaluation variance in GNN benchmarking. Appendix~\ref{app:var_decomp} provides direct regression evidence for this driver: holding the model definition fixed, test accuracy varies significantly ($p<0.05$, $R^2 \in [0.021, 0.289]$) across randomly drawn splits with different test-set homophily, across multiple dataset/model combinations.

Wisconsin and Actor are exceptions, and they are the two datasets identified in Section~\ref{sec:method:hp_priority} as falling outside the regime where \hp{} provides additional value over class-stratified $k$-fold. Both have homophily distributions concentrated in the lowest bins, with the higher bins sparsely populated. The contrast with Amazon-Ratings and Squirrel, which have comparable mean homophily but support across multiple bins, isolates bin skew as the operative factor rather than mean homophily.

Repeating this comparison at \(k=5\) and at \(k=4\) averaged over 6 seeds confirms the result is not an artefact of fold count or seed choice, with \hp{} again achieving the lowest mean rank in both settings (Appendix~\ref{app:acc_per_dataset}). The magnitude of the improvement, not just its direction, follows the same pattern: \hp{} reduces cross-fold standard deviation relative to random \(k\)-fold on a majority of dataset/model combinations across all three settings (Table \ref{tab:stddevdelta} in Appendix~\ref{app:acc_per_dataset}). This variance reduction has a practical consequence, as the identity of the best-performing architecture differs across split strategies on 8 of the 15 datasets, meaning the choice of split protocol can determine which model a benchmark reports as best (Appendix~\ref{app:best_model_shift}).
\section{Conclusion}
\label{sec:conclusion}

We introduced \hp{}, a stratification method for transductive node classification that constructs $k$-fold splits by controlling for both node homophily and class label, extending the stratification principle from the class marginal to the relational structure of the graph~\cite{homophily0}. Evaluated against random $k$-fold and class-stratified $k$-fold across 15 datasets spanning the full homophily spectrum and 7 architectures, \hp{} achieves the lowest mean stability rank, averaged over the 7 models, on 13 of 15 datasets and the best homophily balance across folds on all 15. Class-stratified $k$-fold does not consistently improve over random splitting, indicating that controlling the class marginal alone is insufficient to stabilise GNN evaluation. This advantage holds under alternative fold counts and seeds (Appendix~\ref{app:acc_per_dataset}) and is consistent with a direct link between fold-level homophily and test accuracy across multiple architectures and datasets (Appendix~\ref{app:var_decomp}).

This instability has a direct practical consequence: the identity of the best-performing architecture on a dataset changes with the choice of split strategy on a majority of the datasets evaluated (Appendix~\ref{app:best_model_shift}). Random splitting remains the de facto standard in the literature, and class-stratified splitting, the standard remedy outside graph settings, does not reliably improve on it here. Taken together, these results indicate that split protocol is a substantive source of variance in reported GNN performance, and that accounting for homophily addresses a source of instability that class balance alone does not.

\hp{} is most effective when the homophily distribution has meaningful support across bins. When the distribution is fully degenerate, \hp{} reduces to class-stratified $k$-fold exactly by construction, and the same behaviour emerges gradually under skewed distributions where most nodes concentrate in a small number of bins, as observed on Wisconsin and Actor. This regime is diagnosable from the empirical homophily histogram before any model is trained, and binning strategies that adapt to the empirical density of the distribution, rather than imposing equal-width intervals, are a natural direction for extending \hp{}'s effective regime to these cases. On the remaining datasets spanning the full homophily spectrum, \hp{} improves evaluation stability, indicating that the underlying principle, controlling for the relational structure that governs GNN performance, holds broadly even where the current binning scheme does not.

\section{Acknowledgements}
\label{sec:ack}
The authors gratefully acknowledge the scientific support and HPC resources provided by the Erlangen National High Performance Computing Center (NHR@FAU) of the Friedrich-Alexander-Universität Erlangen-Nürnberg (FAU). The hardware is partially funded by the German Research Foundation (DFG).

\newpage
\bibliographystyle{unsrtnat} 
\bibliography{sections/08_ref}

@ARTICLE{gnnpaper,
  author={Scarselli, Franco and Gori, Marco and Tsoi, Ah Chung and Hagenbuchner, Markus and Monfardini, Gabriele},
  journal={IEEE Transactions on Neural Networks}, 
  title={The Graph Neural Network Model}, 
  year={2009},
  volume={20},
  number={1},
  pages={61-80},
  doi={10.1109/TNN.2008.2005605}}

@article{coradataset,
  title={Collective Classification in Network Data},
  author={Sen, Prithviraj and Namata, Galileo and Bilgic, Mustafa and Getoor, Lise and Gallagher, Brian and Eliassi-Rad, Tina},
  journal={AI Magazine},
  volume={29},
  number={3},
  pages={93--106},
  year={2008},
  publisher={Wiley Online Library}
}

@inproceedings{yangplanetoid,
  title={Revisiting Semi-Supervised Learning with Graph Embeddings},
  author={Yang, Zhilin and Cohen, William W. and Salakhutdinov, Ruslan},
  booktitle={International Conference on Machine Learning},
  year={2016}
}

@inproceedings{platanovchameleondataset,
  title={A Critical Look at the Evaluation of {GNN}s under Heterophily: Are We Really Making Progress?},
  author={Platonov, Oleg and Kuznedelev, Denis and Diskin, Michael and Babenko, Artem and Prokhorenkova, Liudmila},
  booktitle={International Conference on Learning Representations},
  year={2023}
}

@inproceedings{geomgcnpaper,
  title={Geom-{GCN}: Geometric Graph Convolutional Networks},
  author={Pei, Hongbin and Wei, Bingzhe and Chang, Kevin Chen-Chuan and Lei, Yu and Yang, Bo},
  booktitle={International Conference on Learning Representations},
  year={2020}
}

@inproceedings{pitfallspaper,
  title={Pitfalls of Graph Neural Network Evaluation},
  author={Shchur, Oleksandr and Mumme, Maximilian and Bojchevski, Aleksandar and G{\"u}nnemann, Stephan},
  booktitle={Relational Representation Learning Workshop, NeurIPS 2018},
  year={2018}
}

@article{hyptunepitfall,
  title={Classic gnns are strong baselines: Reassessing gnns for node classification},
  author={Luo, Yuankai and Shi, Lei and Wu, Xiao-Ming},
  journal={Advances in Neural Information Processing Systems},
  volume={37},
  pages={97650--97669},
  year={2024}
}

@inproceedings{datasetpitfall,
  title={Position: Graph Learning Will Lose Relevance Due To Poor Benchmarks},
  author={Bechler-Speicher, Maya and Finkelshtein, Ben and Frasca, Fabrizio and M{\"u}ller, Luis and T{\"o}nshoff, Jan and Siraudin, Antoine and Zaverkin, Viktor and Bronstein, Michael M and Niepert, Mathias and Perozzi, Bryan and others},
  booktitle={International Conference on Machine Learning},
  pages={81067--81089},
  year={2025},
  organization={PMLR}
}

@inproceedings{gcnpaper,
  title={Semi-Supervised Classification with Graph Convolutional Networks},
  author={Kipf, Thomas N. and Welling, Max},
  booktitle={International Conference on Learning Representations},
  year={2017}
}

@inproceedings{gatpaper,
  title={Graph Attention Networks},
  author={Veli{\v{c}}kovi{\'c}, Petar and Cucurull, Guillem and Casanova, Arantxa and Romero, Adriana and Li{\`o}, Pietro and Bengio, Yoshua},
  booktitle={International Conference on Learning Representations},
  year={2018}
}

@inproceedings{sagepaper,
  title={Inductive Representation Learning on Large Graphs},
  author={Hamilton, William L. and Ying, Rex and Leskovec, Jure},
  booktitle={Advances in Neural Information Processing Systems},
  year={2017}
}

@inproceedings{appnppaper,
  title={Predict then Propagate: Graph Neural Networks meet Personalized PageRank},
  author={Gasteiger, Johannes and Bojchevski, Aleksandar and G{\"u}nnemann, Stephan},
  booktitle={International Conference on Learning Representations},
  year={2019}
}

@inproceedings{mixhoppaper,
  title={MixHop: Higher-Order Graph Convolutional Architectures via Sparsified Neighborhood Mixing},
  author={Abu-El-Haija, Sami and Perozzi, Bryan and Kapoor, Amol and Alipourfard, Nazanin and Lerman, Kristina and Harutyunyan, Hrayr and Ver Steeg, Greg and Galstyan, Aram},
  booktitle={International Conference on Machine Learning},
  year={2019}
}

@inproceedings{h2gcnpaper,
  title={Beyond Homophily in Graph Neural Networks: Current Limitations and Effective Designs},
  author={Zhu, Jiong and Yan, Yujun and Zhao, Lingxiao and Heimann, Mark and Akoglu, Leman and Koutra, Danai},
  booktitle={Advances in Neural Information Processing Systems},
  year={2020}
}

@inproceedings{gprgnnpaper,
  title={Adaptive Universal Generalized PageRank Graph Neural Network},
  author={Chien, Eli and Peng, Jianhao and Li, Pan and Milenkovic, Olgica},
  booktitle={International Conference on Learning Representations},
  year={2021}
}

@book{kuhnbook,
  title={Applied predictive modeling},
  author={Kuhn, Max and Johnson, Kjell and others},
  volume={26},
  year={2013},
  publisher={Springer},
  pages={67--73}
}

@inproceedings{kohavi1995study,
  title={A Study of Cross-Validation and Bootstrap for Accuracy Estimation and Model Selection},
  author={Kohavi, Ron},
  booktitle={International Joint Conference on Artificial Intelligence},
  year={1995}
}

@article{inductivebiases,
  title={Relational inductive biases, deep learning, and graph networks},
  author={Battaglia, Peter W and Hamrick, Jessica B and Bapst, Victor and Sanchez-Gonzalez, Alvaro and Zambaldi, Vinicius and Malinowski, Mateusz and Tacchetti, Andrea and Raposo, David and Santoro, Adam and Faulkner, Ryan and others},
  journal={arXiv preprint arXiv:1806.01261},
  year={2018}
}

@article{birds,
  title={Birds of a feather: Homophily in social networks},
  author={McPherson, Miller and Smith-Lovin, Lynn and Cook, James M},
  journal={Annual review of sociology},
  volume={27},
  number={1},
  pages={415--444},
  year={2001},
  publisher={Annual Reviews 4139 El Camino Way, PO Box 10139, Palo Alto, CA 94303-0139, USA}
}

@inproceedings{coin,
  title={Two sides of the same coin: Heterophily and oversmoothing in graph convolutional neural networks},
  author={Yan, Yujun and Hashemi, Milad and Swersky, Kevin and Yang, Yaoqing and Koutra, Danai},
  booktitle={2022 IEEE International Conference on Data Mining (ICDM)},
  pages={1287--1292},
  year={2022},
  organization={IEEE}
}

@inproceedings{homophily0,
  title={On performance discrepancies across local homophily levels in graph neural networks},
  author={Loveland, Donald and Zhu, Jiong and Heimann, Mark and Fish, Benjamin and Schaub, Michael T and Koutra, Danai},
  booktitle={Learning on Graphs Conference},
  pages={6--1},
  year={2024},
  organization={PMLR}
}

@inproceedings{homophily1,
  title={Is Homophily a Necessity for Graph Neural Networks?},
  author={Ma, Yao and Liu, Xiaorui and Shah, Neil and Tang, Jiliang},
  booktitle={International Conference on Learning Representations},
  year={2022}
}

@article{homophily3,
  title={Heterophily and graph neural networks: Past, present and future},
  author={Zhu, Jiong and Yan, Yujun and Heimann, Mark and Zhao, Lingxiao and Akoglu, Leman and Koutra, Danai},
  journal={IEEE Data Engineering Bulletin},
  year={2023}
}

@inproceedings{gnnismpnn,
  title={Neural message passing for quantum chemistry},
  author={Gilmer, Justin and Schoenholz, Samuel S and Riley, Patrick F and Vinyals, Oriol and Dahl, George E},
  booktitle={International conference on machine learning},
  pages={1263--1272},
  year={2017},
  organization={PMLR}
}

@article{adjustedhomophily,
  title={Characterizing graph datasets for node classification: Homophily-heterophily dichotomy and beyond},
  author={Platonov, Oleg and Kuznedelev, Denis and Babenko, Artem and Prokhorenkova, Liudmila},
  journal={Advances in Neural Information Processing Systems},
  volume={36},
  pages={523--548},
  year={2023}
}

@article{homophily5,
  title={When do graph neural networks help with node classification? investigating the homophily principle on node distinguishability},
  author={Luan, Sitao and Hua, Chenqing and Xu, Minkai and Lu, Qincheng and Zhu, Jiaqi and Chang, Xiao-Wen and Fu, Jie and Leskovec, Jure and Precup, Doina},
  journal={Advances in Neural Information Processing Systems},
  volume={36},
  pages={28748--28760},
  year={2023}
}

@inproceedings{sechidis2011stratification,
  title={On the Stratification of Multi-Label Data},
  author={Sechidis, Konstantinos and Tsoumakas, Grigorios and Vlahavas, Ioannis},
  booktitle={Joint European Conference on Machine Learning and Knowledge Discovery in Databases},
  year={2011}
}

@article{szymanski2017network,
  title={A Network Perspective on Stratification of Multi-Label Data},
  author={Szyma{\'n}ski, Piotr and Kajdanowicz, Tomasz},
  journal={Proceedings of the First International Workshop on Learning with Imbalanced Domains: Theory and Applications, PMLR},
  year={2017}
}

@article{pyg,
  title={Fast Graph Representation Learning with {PyTorch} Geometric},
  author={Fey, Matthias and Lenssen, Jan E.},
  journal={ICLR Workshop on Representation Learning on Graphs and Manifolds},
  year={2019}
}

@inproceedings{pmbsrs,
  title={On the impact of dataset complexity and sampling strategy in multilabel classifiers performance},
  author={Charte, Francisco and Rivera, Antonio and del Jesus, Mar{\'\i}a Jos{\'e} and Herrera, Francisco},
  booktitle={International conference on hybrid artificial intelligence systems},
  pages={500--511},
  year={2016},
  organization={Springer}
}

@inproceedings{ssstrat,
  title={Stratified sampling for extreme multi-label data},
  author={Merrillees, Maximillian and Du, Lan},
  booktitle={Pacific-Asia Conference on Knowledge Discovery and Data Mining},
  pages={334--345},
  year={2021},
  organization={Springer}
}

@article{optisplit,
  title={Novel split quality measures for stratified multilabel cross validation with application to large and sparse gene ontology datasets},
  author={Tiittanen, Henri and Holm, Liisa and T{\"o}r{\"o}nen, Petri},
  journal={Applied Computing and Intelligence},
  volume={2},
  pages={49--62},
  year={2022}
}

@article{evosplit,
  title={Evosplit: An evolutionary approach to split a multi-label data set into disjoint subsets},
  author={Florez-Revuelta, Francisco},
  journal={Applied Sciences},
  volume={11},
  number={6},
  pages={2823},
  year={2021},
  publisher={MDPI}
}

@inproceedings{jami2025,
  title={Stratify or Die: Rethinking Data Splits in Image Segmentation},
  author={Jami, Naga Venkata Sai Jitin and Altstidl, Thomas and Mueller, Jonas and Li, Jindong and Zanca, Dario and Eskofier, Bjoern and Leutheuser, Heike},
  booktitle={The Thirty-ninth Annual Conference on Neural Information Processing Systems},
  year={2025},
  url={https://openreview.net/pdf?id=ngtFOxkQ8b}
}

@inproceedings{axdependence,
  title={Feature Distribution on Graph Topology Mediates the Effect of Graph Convolution: Homophily Perspective},
  author={Lee, Soo Yong and Kim, Sunwoo and Bu, Fanchen and Yoo, Jaemin and Tang, Jiliang and Shin, Kijung},
  booktitle={International Conference on Machine Learning},
  pages={26686--26714},
  year={2024},
  organization={PMLR}
}

@article{trihom,
  title={What is missing for graph homophily? disentangling graph homophily for graph neural networks},
  author={Zheng, Yilun and Luan, Sitao and Chen, Lihui},
  journal={Advances in Neural Information Processing Systems},
  volume={37},
  pages={68406--68452},
  year={2024}
}

@article{linkx,
  title={Large scale learning on non-homophilous graphs: New benchmarks and strong simple methods},
  author={Lim, Derek and Hohne, Felix and Li, Xiuyu and Huang, Sijia Linda and Gupta, Vaishnavi and Bhalerao, Omkar and Lim, Ser Nam},
  journal={Advances in neural information processing systems},
  volume={34},
  pages={20887--20902},
  year={2021}
}

@inproceedings{monetpaper,
  title={Geometric deep learning on graphs and manifolds using mixture model cnns},
  author={Monti, Federico and Boscaini, Davide and Masci, Jonathan and Rodola, Emanuele and Svoboda, Jan and Bronstein, Michael M},
  booktitle={Proceedings of the IEEE conference on computer vision and pattern recognition},
  pages={5115--5124},
  year={2017}
}

\newpage
\appendix

\numberwithin{table}{section}
\numberwithin{figure}{section}
\numberwithin{algorithm}{section}

\section{\hp{} Algorithm}
\label{app:algorithm}

Algorithm~\ref{alg:hp_priority} provides the full pseudocode for \hp{}, complementing the prose description in Section~\ref{sec:method:hp_priority}.

\begin{algorithm}[h]
\caption{\hp{}: \textbf{H}omophily-\textbf{P}riority \textbf{Strat}ification}
\label{alg:hp_priority}
\hspace*{\algorithmicindent} \textbf{Input}: Graph \(\mathcal{G} = (\mathcal{V}, \mathcal{E})\), node labels \(\mathbf{Y}\), number of folds \(k\), number of bins \(B\).\\
\hspace*{\algorithmicindent} \textbf{Output}: \(k\)-fold partition \(\mathbf{S} = \{S_1, \ldots, S_k\}\) of \(\mathcal{V}\)
\begin{algorithmic}[1]
\State Compute node homophily \(h_v\) for all \(v \in \mathcal{V}\) \hfill (Eq.~\ref{eq:node_homophily})
\State Partition nodeset \(\mathcal{V}\) into \(B\) equal-width bins \(\mathcal{V}_1, \ldots, \mathcal{V}_B\) based on \(h_v\)
\State Initialise subset \(|S_i| \leftarrow 0\) for all \(i \in \{1, \ldots, k\}\)
\State \texttt{\# Loop through the bins}
\For{each bin \(b \in \{1, \ldots, B\}\)}
    \State \texttt{\# Loop through the classes}
    \For{each class \(c \in \{1, \ldots, C\}\)}
        \State \texttt{\# Shuffle the nodes in a stratum}
        \State \(\tilde{\mathcal{V}}_{b,c} \leftarrow \text{shuffle}(\mathcal{V}_{b,c})\)
        \State \texttt{\# Choose subset with fewest nodes; ties broken by lowest index}
        \State \(i^* \leftarrow \min\bigl\{i : |S_i| = \min_{j}\, |S_j|\bigr\}\)
        \State \texttt{\# Assign nodes to subsets in round-robin order}
        \For{each \(v_b^c \in \tilde{\mathcal{V}}_{b,c}\)}
            \State \texttt{\# Assign nodes to \(i^*\) subset}
            \State \(S_{i^*} \leftarrow S_{i^*} \cup \{v_b^c\}\)
            \State \texttt{\# Reset subset of choice for next node in stratum}
            \State \(i^* \leftarrow (i^* \bmod k) + 1\)
        \EndFor
    \EndFor
\EndFor
\end{algorithmic}
\end{algorithm}

\newpage
\section{Variance Decomposition under Random \texorpdfstring{$k$}{k}-fold}
\label{app:var_decomp}

The motivation for \hp{} rests on the claim that fold-level homophily composition is a substantive driver of cross-fold accuracy variance under random splits. To test this directly, we ran \(50\) seeds of random \(k\)-fold for \(9\) model/dataset combinations spanning homophily-assuming and heterophily-robust architectures and the full homophily spectrum, yielding \(200\) folds per combination. For each fold, we recorded the mean test-fold homophily and the test accuracy, and regressed accuracy on test-fold homophily within each combination.

\begin{figure}[h]
\centering
\includegraphics[width=0.8\textwidth]{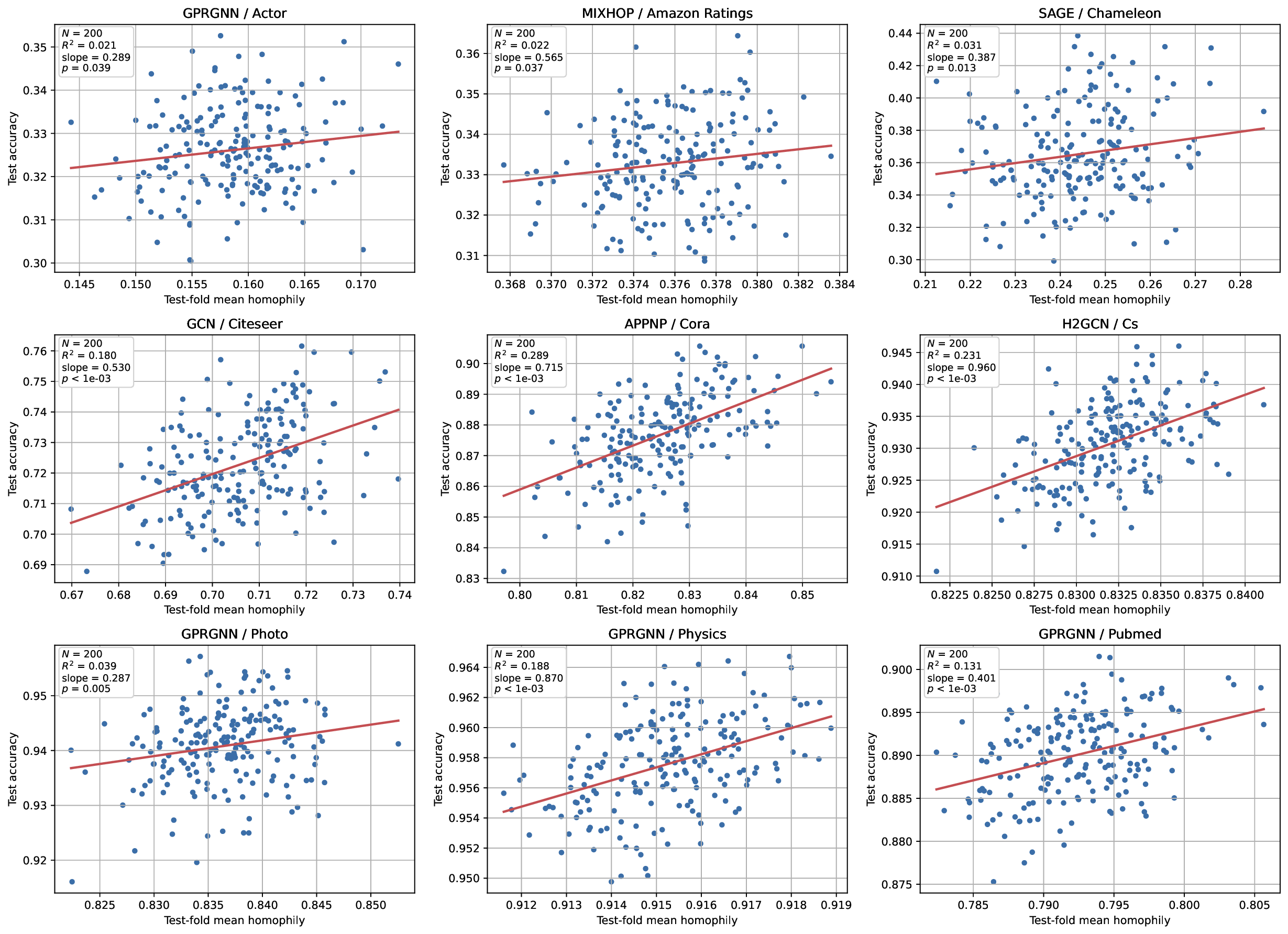}
\caption{Fold-level test accuracy versus fold-level mean homophily under random \(k\)-fold splits, for the \(9\) model/dataset combinations in Table~\ref{tab:var_decomp1} (\(N=200\) folds per panel, 50 seeds). Each panel reports its fitted regression line and corresponding \(R^2\), \(p\)-value, and slope.}
\label{fig:var_decomp}
\end{figure}

Table~\ref{tab:var_decomp1} reports the regression statistics for all \(9\) combinations. Fold-level homophily is a statistically significant predictor of test accuracy in every combination tested (\(p < 0.05\)), with \(R^2\) ranging from \(0.021\) on GPR-GNN/Actor to \(0.289\) on APPNP/Cora. This generalises the association reported in the main text beyond a single model/dataset pair, consistent with the theoretical predictions of~\cite{homophily0} on local homophily mismatch: fold-level homophily composition is a systematic source of accuracy variance under random splitting, not an artefact specific to one model or dataset. Figure~\ref{fig:var_decomp} shows the fitted regression for each combination.

\begin{table}[h]
\centering
\caption{Regression of fold-level test accuracy on fold-level mean homophily under random \(k\)-fold, for 9 model/dataset combinations (\(N=200\) folds per combination, 50 seeds). All combinations are statistically significant at \(p<0.05\).}
\label{tab:var_decomp1}
\begin{tabular}{llcccc}
\toprule
Model & Dataset & \(N\) & \(R^2\) & \(p\)-value & Slope \\
\midrule
APPNP    & Cora           & 200 & 0.289 & \(2.28\times10^{-16}\) & 0.715 \\
H2GCN    & CS             & 200 & 0.231 & \(6.15\times10^{-13}\) & 0.960 \\
GPR-GNN  & Physics        & 200 & 0.188 & \(1.51\times10^{-10}\) & 0.870 \\
GCN      & CiteSeer       & 200 & 0.180 & \(3.63\times10^{-10}\) & 0.530 \\
GPR-GNN  & PubMed         & 200 & 0.131 & \(1.36\times10^{-07}\) & 0.401 \\
GPR-GNN  & Photo          & 200 & 0.039 & \(5.34\times10^{-03}\) & 0.287 \\
GraphSAGE & Chameleon     & 200 & 0.031 & \(1.28\times10^{-02}\) & 0.387 \\
MixHop   & Amazon-Ratings & 200 & 0.022 & \(3.69\times10^{-02}\) & 0.565 \\
GPR-GNN  & Actor          & 200 & 0.021 & \(3.88\times10^{-02}\) & 0.289 \\
\bottomrule
\end{tabular}
\end{table}

\section{Per-Dataset Accuracy Results}
\label{app:acc_per_dataset}

Tables~\ref{tab:acc_per_dataset_k4}, \ref{tab:acc_per_dataset_k5}, and \ref{tab:acc_per_dataset_k4_seeds} report mean test accuracy and cross-fold standard deviation for all 15 datasets, 7 models, and 3 split strategies, under \(k=4\), \(k=5\), and \(k=4\) averaged over 6 random seeds (42, 0, 1, 2, 3, 4; 24 runs total), respectively. Mean accuracy is reported as a percentage, and standard deviation is multiplied by \(10^2\) for readability. Bold values indicate the lowest standard deviation per (dataset, model) pair. The Rank column in each table reports the average rank of each strategy across all 7 models for that dataset, and the Avg.\ Rank rows at the bottom report the average rank of each strategy across all 15 datasets for each model. Across all three settings, \hp{} achieves the lowest overall mean rank of the three strategies (\(k=4\): 1.49 vs.\ 2.31 for random \(k\)-fold; \(k=5\): 1.63 vs.\ 2.11; \(k=4\), 6 seeds: 1.53 vs.\ 2.34), confirming that the stability advantage reported in Section~\ref{sec:results:accuracy} is not an artefact of the specific fold count or seed used in the main results.

\begin{figure}[h]
\centering
\includegraphics[width=\textwidth]{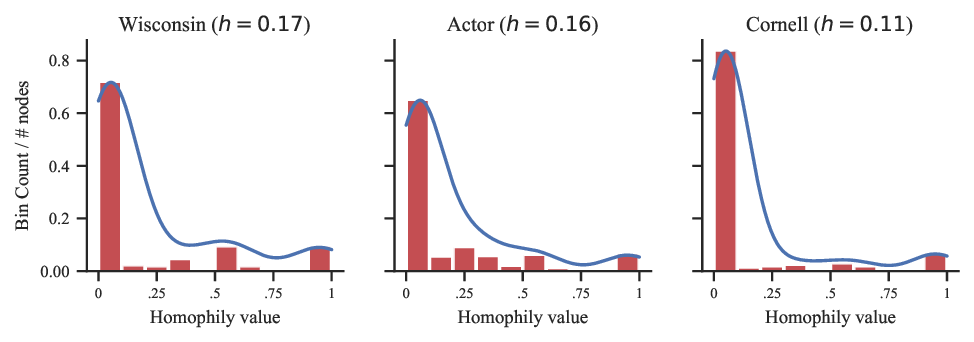}
\caption{Node homophily distributions for the three datasets on which \hp{} fails to achieve the lowest stability rank in at least one of the \(k=4\), \(k=5\), or 6-seed \(k=4\) settings, binned over $[0, 1]$. Wisconsin ($\bar{h} = 0.17$) and Actor ($\bar{h} = 0.16$) concentrate most nodes in the lowest bins, leaving higher bins sparsely populated. Cornell ($\bar{h} = 0.11$) shows a similar concentration, reducing the number of nodes available per bin once folds are drawn at smaller test-set sizes. In all three datasets, the homophily axis carries little discriminative information for stratification, consistent with the failure mode discussed in Section~\ref{sec:method:hp_priority}.}
\label{fig:homophily_distributions_appendix}
\end{figure}

The datasets on which \hp{} fails to achieve the lowest rank are consistent with the failure mode identified in Section~\ref{sec:method:hp_priority}: datasets whose homophily distribution is skewed into a small number of bins. At \(k=4\), these are Wisconsin and Actor. At \(k=5\), the smaller per-fold test sets exaggerate this effect and the set widens to Wisconsin, Actor, and Cornell. Figure~\ref{fig:homophily_distributions_appendix} shows the homophily distributions for these datasets, illustrating the degree of bin skew responsible for this behaviour.

Table~\ref{tab:stddevdelta} reports the reduction in cross-fold standard deviation of test accuracy relative to random \(k\)-fold, for class-stratified \(k\)-fold and \hp{}, under the same three settings. For each dataset and model, the delta is the standard deviation under random \(k\)-fold minus the standard deviation under the alternative strategy: positive means lower variance than random, negative means higher.

\newpage

\begin{table}[h]
\centering
\caption{Mean test accuracy (\%) and cross-fold standard deviation (\(\times 10^2\)) at \(k=4\) for all datasets, models, and split strategies. \textbf{Bold} std = lowest per (dataset, model). Ratings = Amazon-Ratings; Roman = Roman-Empire.}
\label{tab:acc_per_dataset_k4}
\renewcommand{\arraystretch}{0.85}
\scriptsize
\resizebox{\textwidth}{!}{%
\begin{tabular}{ll|cc|cc|cc|cc|cc|cc|cc|c}
\toprule
 & & \multicolumn{2}{c|}{GCN} & \multicolumn{2}{c|}{GAT} & \multicolumn{2}{c|}{GraphSAGE} & \multicolumn{2}{c|}{APPNP} & \multicolumn{2}{c|}{MixHop} & \multicolumn{2}{c|}{H2GCN} & \multicolumn{2}{c|}{GPR-GNN} & \\
Dataset & Strategy & Mean & Std & Mean & Std & Mean & Std & Mean & Std & Mean & Std & Mean & Std & Mean & Std & Rank \\
\midrule
\multirow{3}{*}{Physics} & Rand. & 95.17 & 0.44 & 94.67 & 0.48 & 95.57 & 0.26 & 94.99 & 0.40 & 95.36 & 0.34 & 95.62 & 0.22 & 95.65 & 0.34 & 2.86 \\
 & Class. & 95.08 & 0.27 & 94.41 & \textbf{0.16} & 95.43 & \textbf{0.06} & 94.89 & 0.32 & 95.34 & \textbf{0.15} & 95.60 & 0.27 & 95.68 & 0.29 & 1.71 \\
 & \hp{} & 95.18 & \textbf{0.14} & 94.56 & 0.33 & 95.57 & 0.20 & 94.84 & \textbf{0.19} & 95.44 & 0.22 & 95.52 & \textbf{0.18} & 95.72 & \textbf{0.26} & \textbf{1.43} \\
\hline
\multirow{3}{*}{Photo} & Rand. & 92.76 & 1.03 & 93.64 & 0.77 & 93.04 & \textbf{0.40} & 93.60 & 1.16 & 93.95 & 0.91 & 92.42 & 0.91 & 93.76 & 0.86 & 1.86 \\
 & Class. & 92.71 & 1.49 & 93.64 & 1.37 & 93.56 & 1.15 & 93.41 & 1.41 & 93.78 & 1.00 & 92.92 & 0.97 & 93.79 & 1.39 & 3.00 \\
 & \hp{} & 92.81 & \textbf{0.41} & 93.74 & \textbf{0.26} & 93.50 & 0.49 & 93.44 & \textbf{0.31} & 93.88 & \textbf{0.15} & 93.31 & \textbf{0.11} & 94.15 & \textbf{0.32} & \textbf{1.14} \\
\hline
\multirow{3}{*}{CS} & Rand. & 91.38 & 0.40 & 90.74 & 0.72 & 91.94 & 0.68 & 91.77 & 0.71 & 92.28 & 0.40 & 93.38 & 0.60 & 93.05 & 0.75 & 2.14 \\
 & Class. & 91.40 & 0.92 & 90.58 & 1.08 & 91.50 & 1.03 & 91.64 & 0.85 & 91.95 & 0.83 & 93.19 & 0.51 & 92.68 & 0.79 & 2.86 \\
 & \hp{} & 91.35 & \textbf{0.24} & 90.47 & \textbf{0.49} & 91.84 & \textbf{0.54} & 91.74 & \textbf{0.36} & 92.11 & \textbf{0.24} & 93.03 & \textbf{0.45} & 92.95 & \textbf{0.42} & \textbf{1.00} \\
\hline
\multirow{3}{*}{Cora} & Rand. & 85.90 & 2.37 & 86.29 & 2.01 & 85.14 & 2.26 & 86.95 & 1.72 & 85.50 & 1.61 & 85.35 & 2.88 & 85.96 & 2.04 & 2.86 \\
 & Class. & 86.59 & \textbf{1.19} & 86.60 & 1.66 & 86.05 & 1.84 & 88.19 & 1.59 & 86.44 & 1.94 & 85.87 & 1.59 & 87.15 & 1.69 & 2.00 \\
 & \hp{} & 86.50 & 1.42 & 86.62 & \textbf{0.49} & 86.28 & \textbf{1.25} & 87.66 & \textbf{0.74} & 85.82 & \textbf{1.12} & 84.99 & \textbf{0.43} & 87.29 & \textbf{0.96} & \textbf{1.14} \\
\hline
\multirow{3}{*}{PubMed} & Rand. & 86.87 & 0.49 & 85.33 & 1.11 & 88.28 & 0.75 & 86.50 & 0.77 & 88.73 & 0.79 & 89.03 & 0.55 & 89.04 & 0.33 & 2.71 \\
 & Class. & 86.82 & \textbf{0.25} & 85.25 & 0.39 & 88.09 & \textbf{0.34} & 86.47 & 0.54 & 88.67 & \textbf{0.20} & 89.03 & 0.64 & 88.92 & 0.65 & 1.86 \\
 & \hp{} & 86.83 & 0.31 & 85.48 & \textbf{0.38} & 88.25 & 0.42 & 86.31 & \textbf{0.39} & 88.64 & 0.30 & 88.94 & \textbf{0.53} & 89.10 & \textbf{0.31} & \textbf{1.43} \\
\hline
\multirow{3}{*}{Computers} & Rand. & 88.49 & 1.00 & 90.37 & 0.70 & 87.11 & 0.95 & 88.36 & \textbf{0.37} & 91.79 & \textbf{0.38} & 89.88 & 0.70 & 90.32 & 0.87 & 2.29 \\
 & Class. & 88.89 & 0.78 & 89.91 & \textbf{0.53} & 87.21 & 1.47 & 87.68 & 1.06 & 89.39 & 1.59 & 90.09 & 0.62 & 89.77 & 0.81 & 2.14 \\
 & \hp{} & 88.47 & \textbf{0.45} & 90.71 & 0.69 & 88.12 & \textbf{0.67} & 88.07 & 0.95 & 90.76 & 1.77 & 90.21 & \textbf{0.51} & 90.65 & \textbf{0.41} & \textbf{1.57} \\
\hline
\multirow{3}{*}{CiteSeer} & Rand. & 72.14 & 1.39 & 71.69 & 1.51 & 70.46 & 1.74 & 71.21 & 1.76 & 70.09 & 2.06 & 68.72 & 2.02 & 71.08 & 2.12 & 2.71 \\
 & Class. & 72.24 & \textbf{1.06} & 72.02 & \textbf{1.28} & 71.14 & 1.97 & 71.52 & \textbf{1.00} & 70.64 & 1.76 & 69.14 & 1.84 & 71.62 & 1.78 & 1.71 \\
 & \hp{} & 73.19 & 1.92 & 72.60 & 1.37 & 71.80 & \textbf{0.69} & 72.31 & 1.45 & 70.92 & \textbf{0.88} & 70.16 & \textbf{1.55} & 71.69 & \textbf{1.20} & \textbf{1.57} \\
\hline
\multirow{3}{*}{Ratings} & Rand. & 27.06 & 0.35 & 25.38 & \textbf{0.33} & 29.48 & 1.12 & 26.29 & 0.34 & 33.12 & \textbf{0.91} & 31.14 & 1.12 & 30.93 & 0.84 & 2.14 \\
 & Class. & 26.71 & 0.78 & 25.61 & 0.51 & 29.70 & 0.65 & 26.15 & 0.30 & 33.70 & 1.33 & 31.33 & 1.44 & 30.64 & \textbf{0.48} & 2.29 \\
 & \hp{} & 27.11 & \textbf{0.30} & 25.42 & 0.40 & 29.79 & \textbf{0.35} & 26.47 & \textbf{0.13} & 33.33 & 1.45 & 31.84 & \textbf{0.97} & 30.94 & 0.59 & \textbf{1.57} \\
\hline
\multirow{3}{*}{Chameleon} & Rand. & 35.89 & 3.61 & 33.53 & 1.74 & 35.77 & 1.71 & 34.77 & 2.69 & 33.87 & \textbf{0.36} & 30.51 & 5.19 & 32.62 & 4.56 & 2.14 \\
 & Class. & 35.93 & 2.40 & 36.24 & 1.91 & 36.49 & 1.89 & 34.42 & 2.77 & 37.06 & 4.67 & 36.87 & 6.28 & 33.49 & \textbf{0.73} & 2.57 \\
 & \hp{} & 35.83 & \textbf{1.25} & 34.76 & \textbf{1.24} & 36.77 & \textbf{1.18} & 34.83 & \textbf{1.08} & 34.77 & 1.58 & 37.59 & \textbf{4.34} & 35.22 & 2.24 & \textbf{1.29} \\
\hline
\multirow{3}{*}{Squirrel} & Rand. & 27.41 & 2.50 & 24.17 & 2.24 & 28.62 & 1.73 & 27.27 & 1.27 & 26.12 & 2.73 & 30.71 & 1.52 & 27.40 & \textbf{1.11} & 2.71 \\
 & Class. & 25.29 & 1.13 & 23.07 & \textbf{1.52} & 28.37 & 1.44 & 24.39 & 1.30 & 26.69 & \textbf{1.38} & 29.25 & \textbf{1.37} & 25.13 & 2.10 & 1.86 \\
 & \hp{} & 24.00 & \textbf{1.11} & 23.72 & 2.19 & 27.89 & \textbf{1.33} & 24.75 & \textbf{1.19} & 25.84 & 2.68 & 29.26 & \textbf{1.37} & 25.29 & \textbf{1.11} & \textbf{1.43} \\
\hline
\multirow{3}{*}{Wisconsin} & Rand. & 27.20 & \textbf{4.13} & 25.19 & 5.28 & 56.05 & 3.05 & 26.93 & \textbf{4.62} & 43.58 & 12.01 & 59.47 & \textbf{6.39} & 45.86 & 18.92 & \textbf{1.86} \\
 & Class. & 28.57 & 4.78 & 28.85 & 5.50 & 50.36 & \textbf{1.11} & 31.39 & 5.40 & 45.28 & 10.77 & 62.61 & 8.21 & 43.70 & \textbf{16.21} & \textbf{1.86} \\
 & \hp{} & 27.33 & 4.91 & 30.40 & \textbf{4.03} & 55.54 & 7.78 & 29.27 & 6.28 & 39.85 & \textbf{8.02} & 59.49 & 8.38 & 43.12 & 16.47 & 2.29 \\
\hline
\multirow{3}{*}{Actor} & Rand. & 23.68 & 0.95 & 23.49 & 1.34 & 30.04 & 0.77 & 23.66 & \textbf{0.66} & 32.12 & \textbf{0.41} & 30.39 & 1.82 & 33.12 & 1.57 & 2.14 \\
 & Class. & 23.74 & \textbf{0.79} & 23.97 & \textbf{0.75} & 29.89 & 0.91 & 24.07 & 1.00 & 32.53 & 0.67 & 29.75 & 1.73 & 33.30 & 1.14 & \textbf{1.86} \\
 & \hp{} & 24.00 & 1.29 & 23.78 & 0.92 & 29.88 & \textbf{0.53} & 23.92 & 1.73 & 31.38 & 0.92 & 31.10 & \textbf{1.43} & 32.64 & \textbf{0.65} & 2.00 \\
\hline
\multirow{3}{*}{Cornell} & Rand. & 22.27 & 3.50 & 22.41 & 4.24 & 50.11 & 12.25 & 22.01 & 4.79 & 35.17 & 11.89 & 57.84 & 12.89 & 23.43 & 7.62 & 2.57 \\
 & Class. & 24.09 & \textbf{2.39} & 20.42 & \textbf{2.44} & 50.02 & 15.81 & 22.92 & 3.44 & 44.12 & 14.22 & 58.55 & 4.50 & 21.31 & 2.06 & 2.00 \\
 & \hp{} & 26.12 & 5.98 & 23.23 & 3.49 & 59.43 & \textbf{11.07} & 21.21 & \textbf{1.90} & 37.92 & \textbf{6.44} & 58.25 & \textbf{2.80} & 22.48 & \textbf{1.83} & \textbf{1.43} \\
\hline
\multirow{3}{*}{Texas} & Rand. & 28.74 & \textbf{4.23} & 23.30 & \textbf{2.00} & 67.67 & 8.44 & 26.41 & 4.00 & 33.56 & 7.35 & 65.56 & 14.32 & 36.84 & 24.48 & 1.86 \\
 & Class. & 28.57 & 6.81 & 28.50 & 10.59 & 56.55 & 8.92 & 26.80 & 8.51 & 37.67 & 11.10 & 65.52 & 14.63 & 28.96 & \textbf{7.76} & 2.71 \\
 & \hp{} & 27.17 & 5.86 & 26.78 & 7.30 & 62.18 & \textbf{7.03} & 24.75 & \textbf{3.69} & 32.39 & \textbf{2.44} & 70.69 & \textbf{13.72} & 35.62 & 22.19 & \textbf{1.43} \\
\hline
\multirow{3}{*}{Roman} & Rand. & 38.06 & \textbf{0.42} & 28.84 & 0.68 & 64.33 & 0.54 & 37.61 & 0.52 & 63.37 & \textbf{0.39} & 70.89 & \textbf{0.13} & 58.28 & 1.17 & 1.86 \\
 & Class. & 38.10 & 0.56 & 29.04 & 0.44 & 64.30 & 1.08 & 37.75 & 0.82 & 63.18 & 1.58 & 71.52 & 0.98 & 58.40 & \textbf{0.19} & 2.57 \\
 & \hp{} & 38.32 & 0.45 & 28.94 & \textbf{0.41} & 64.54 & \textbf{0.49} & 38.03 & \textbf{0.31} & 63.02 & 0.55 & 71.63 & 0.58 & 58.24 & 1.04 & \textbf{1.57} \\
\midrule
\multirow{3}{*}{Avg.\ Rank} & Rand. & \multicolumn{2}{c|}{2.20} & \multicolumn{2}{c|}{2.47} & \multicolumn{2}{c|}{2.27} & \multicolumn{2}{c|}{2.20} & \multicolumn{2}{c|}{2.00} & \multicolumn{2}{c|}{2.33} & \multicolumn{2}{c|}{2.73} & 2.31 \\
 & Class. & \multicolumn{2}{c|}{2.00} & \multicolumn{2}{c|}{2.00} & \multicolumn{2}{c|}{2.40} & \multicolumn{2}{c|}{2.40} & \multicolumn{2}{c|}{2.27} & \multicolumn{2}{c|}{2.40} & \multicolumn{2}{c|}{1.93} & 2.20 \\
 & \hp{} & \multicolumn{2}{c|}{\textbf{1.80}} & \multicolumn{2}{c|}{\textbf{1.53}} & \multicolumn{2}{c|}{\textbf{1.33}} & \multicolumn{2}{c|}{\textbf{1.40}} & \multicolumn{2}{c|}{\textbf{1.73}} & \multicolumn{2}{c|}{\textbf{1.27}} & \multicolumn{2}{c|}{\textbf{1.33}} & \textbf{1.49} \\
\bottomrule
\end{tabular}}
\end{table}

\newpage

\begin{table}[h]
\centering
\caption{Mean test accuracy (\%) and cross-fold standard deviation (\(\times 10^2\)) at \(k=5\) for all datasets, models, and split strategies. \textbf{Bold} std = lowest per (dataset, model). Ratings = Amazon-Ratings; Roman = Roman-Empire.}
\label{tab:acc_per_dataset_k5}
\renewcommand{\arraystretch}{0.85}
\scriptsize
\resizebox{\textwidth}{!}{%
\begin{tabular}{ll|cc|cc|cc|cc|cc|cc|cc|c}
\toprule
 & & \multicolumn{2}{c|}{GCN} & \multicolumn{2}{c|}{GAT} & \multicolumn{2}{c|}{GraphSAGE} & \multicolumn{2}{c|}{APPNP} & \multicolumn{2}{c|}{MixHop} & \multicolumn{2}{c|}{H2GCN} & \multicolumn{2}{c|}{GPR-GNN} & \\
Dataset & Strategy & Mean & Std & Mean & Std & Mean & Std & Mean & Std & Mean & Std & Mean & Std & Mean & Std & Rank \\
\midrule
\multirow{3}{*}{Physics} & Rand. & 95.19 & 0.30 & 94.74 & 0.43 & 95.62 & \textbf{0.16} & 95.04 & 0.34 & 95.62 & 0.28 & 95.68 & \textbf{0.11} & 95.94 & 0.21 & 2.14 \\
 & Class. & 95.17 & 0.33 & 94.59 & 0.37 & 95.52 & 0.33 & 94.83 & \textbf{0.31} & 95.56 & \textbf{0.08} & 95.67 & 0.27 & 95.81 & 0.28 & 2.29 \\
 & \hp{} & 95.19 & \textbf{0.27} & 94.68 & \textbf{0.29} & 95.62 & 0.29 & 94.99 & 0.32 & 95.76 & 0.16 & 95.73 & 0.17 & 95.84 & \textbf{0.12} & \textbf{1.57} \\
\hline
\multirow{3}{*}{Photo} & Rand. & 92.39 & 1.42 & 93.89 & 0.88 & 93.32 & 1.36 & 93.11 & 1.35 & 94.08 & 0.58 & 93.29 & 0.84 & 94.00 & 0.70 & 2.14 \\
 & Class. & 92.50 & 1.56 & 93.80 & 1.47 & 93.60 & 1.62 & 93.65 & 1.22 & 93.97 & 1.25 & 93.27 & 1.60 & 94.15 & 1.19 & 2.86 \\
 & \hp{} & 92.40 & \textbf{0.67} & 93.60 & \textbf{0.50} & 93.01 & \textbf{0.73} & 93.19 & \textbf{0.75} & 93.64 & \textbf{0.56} & 93.41 & \textbf{0.45} & 93.79 & \textbf{0.33} & \textbf{1.00} \\
\hline
\multirow{3}{*}{CS} & Rand. & 91.59 & 0.71 & 90.70 & 0.83 & 92.21 & \textbf{0.46} & 91.78 & 0.65 & 92.33 & 0.75 & 93.38 & \textbf{0.32} & 93.21 & 0.99 & 2.00 \\
 & Class. & 91.46 & 0.83 & 90.69 & 1.03 & 92.07 & 0.77 & 91.64 & 0.82 & 92.10 & 0.74 & 93.32 & 0.69 & 92.84 & \textbf{0.60} & 2.43 \\
 & \hp{} & 91.48 & \textbf{0.36} & 90.83 & \textbf{0.51} & 92.10 & 0.66 & 91.91 & \textbf{0.51} & 92.20 & \textbf{0.64} & 93.28 & 0.80 & 93.13 & 0.86 & \textbf{1.57} \\
\hline
\multirow{3}{*}{Cora} & Rand. & 86.25 & 1.34 & 86.80 & 1.22 & 86.06 & \textbf{1.79} & 88.29 & 1.08 & 86.49 & 1.42 & 85.70 & 2.76 & 87.43 & 1.35 & 2.14 \\
 & Class. & 87.53 & 2.92 & 87.18 & \textbf{1.11} & 86.51 & 1.92 & 88.32 & 1.22 & 86.68 & 1.49 & 85.94 & 2.15 & 87.38 & 1.30 & 2.29 \\
 & \hp{} & 87.19 & \textbf{1.23} & 87.20 & 1.69 & 85.56 & 2.36 & 87.95 & \textbf{0.92} & 86.33 & \textbf{1.29} & 85.97 & \textbf{1.40} & 87.09 & \textbf{1.22} & \textbf{1.57} \\
\hline
\multirow{3}{*}{PubMed} & Rand. & 86.85 & 0.88 & 85.27 & 1.02 & 88.41 & 0.77 & 86.49 & 0.79 & 89.34 & 0.63 & 89.41 & 0.64 & 89.17 & 0.79 & 2.43 \\
 & Class. & 86.92 & 1.00 & 85.33 & 0.49 & 88.28 & 0.94 & 86.30 & 0.76 & 88.94 & 0.58 & 89.00 & 0.99 & 89.32 & 0.84 & 2.57 \\
 & \hp{} & 86.89 & \textbf{0.35} & 85.49 & \textbf{0.20} & 88.24 & \textbf{0.23} & 86.56 & \textbf{0.40} & 89.13 & \textbf{0.40} & 89.48 & \textbf{0.43} & 89.16 & \textbf{0.53} & \textbf{1.00} \\
\hline
\multirow{3}{*}{Computers} & Rand. & 88.89 & \textbf{0.47} & 90.39 & 0.83 & 86.83 & 2.28 & 88.05 & 0.95 & 91.34 & \textbf{0.98} & 89.93 & 0.57 & 90.19 & 0.84 & 1.86 \\
 & Class. & 88.78 & 0.96 & 90.33 & 1.04 & 86.55 & 1.75 & 88.34 & 1.26 & 91.11 & 1.92 & 90.60 & 0.63 & 90.66 & 1.05 & 2.86 \\
 & \hp{} & 88.84 & 0.58 & 90.42 & \textbf{0.27} & 87.64 & \textbf{1.12} & 88.01 & \textbf{0.84} & 90.66 & 1.13 & 90.68 & \textbf{0.30} & 89.85 & \textbf{0.71} & \textbf{1.29} \\
\hline
\multirow{3}{*}{CiteSeer} & Rand. & 72.63 & 1.93 & 73.15 & 2.18 & 71.21 & 2.16 & 72.47 & 1.45 & 71.35 & 2.26 & 69.96 & 1.44 & 71.76 & 2.13 & 2.71 \\
 & Class. & 73.11 & \textbf{1.04} & 72.34 & \textbf{0.83} & 71.91 & 1.59 & 71.82 & \textbf{0.94} & 71.10 & 1.71 & 70.66 & 1.79 & 71.83 & 1.61 & 1.71 \\
 & \hp{} & 73.38 & 1.49 & 72.51 & 0.96 & 70.82 & \textbf{1.12} & 72.33 & 1.56 & 70.85 & \textbf{0.97} & 69.91 & \textbf{1.41} & 71.55 & \textbf{1.20} & \textbf{1.57} \\
\hline
\multirow{3}{*}{Ratings} & Rand. & 27.15 & \textbf{0.30} & 25.35 & 0.36 & 29.25 & 0.84 & 26.11 & \textbf{0.31} & 33.66 & \textbf{0.62} & 31.14 & 1.26 & 30.87 & \textbf{0.48} & \textbf{1.71} \\
 & Class. & 26.89 & 0.59 & 25.30 & 0.57 & 29.40 & 0.43 & 26.21 & 0.35 & 34.14 & 1.12 & 30.40 & \textbf{1.03} & 30.53 & 0.86 & 2.43 \\
 & \hp{} & 27.11 & 0.44 & 25.33 & \textbf{0.23} & 29.11 & \textbf{0.20} & 26.30 & 0.52 & 34.42 & 0.79 & 31.44 & 1.13 & 30.97 & 0.74 & 1.86 \\
\hline
\multirow{3}{*}{Chameleon} & Rand. & 35.21 & 3.68 & 33.88 & 1.29 & 37.74 & 3.91 & 34.24 & 3.46 & 34.74 & 1.97 & 32.13 & 6.63 & 34.31 & 2.91 & 2.43 \\
 & Class. & 35.62 & \textbf{3.12} & 35.85 & 1.70 & 36.39 & \textbf{3.18} & 35.01 & 2.54 & 38.53 & 2.82 & 35.58 & \textbf{4.37} & 34.85 & 3.06 & 2.00 \\
 & \hp{} & 38.45 & 3.85 & 35.39 & \textbf{1.11} & 36.05 & 3.30 & 36.02 & \textbf{1.99} & 37.04 & \textbf{1.54} & 35.96 & 6.18 & 36.03 & \textbf{2.44} & \textbf{1.57} \\
\hline
\multirow{3}{*}{Squirrel} & Rand. & 26.99 & 2.13 & 24.66 & \textbf{1.65} & 29.33 & 2.77 & 27.10 & 1.86 & 25.69 & \textbf{1.28} & 30.40 & 4.25 & 27.51 & 2.73 & 2.29 \\
 & Class. & 27.01 & 2.79 & 23.41 & 2.13 & 28.65 & \textbf{1.79} & 26.52 & 1.62 & 26.20 & 1.47 & 28.32 & \textbf{1.80} & 24.48 & 2.64 & 2.14 \\
 & \hp{} & 24.94 & \textbf{1.13} & 23.47 & 1.82 & 29.41 & 2.05 & 26.51 & \textbf{1.23} & 25.20 & 1.40 & 30.44 & 2.42 & 26.32 & \textbf{1.01} & \textbf{1.57} \\
\hline
\multirow{3}{*}{Wisconsin} & Rand. & 27.60 & 5.05 & 26.01 & \textbf{5.83} & 54.98 & \textbf{7.59} & 28.51 & 5.89 & 48.92 & 13.96 & 60.91 & \textbf{6.68} & 50.79 & 17.23 & \textbf{1.86} \\
 & Class. & 27.42 & 6.03 & 29.75 & 9.82 & 55.00 & 8.17 & 27.39 & \textbf{1.19} & 49.53 & 15.89 & 66.74 & 10.24 & 60.44 & 15.08 & 2.14 \\
 & \hp{} & 28.87 & \textbf{4.11} & 31.15 & 10.17 & 64.36 & 16.59 & 28.14 & 2.89 & 36.58 & \textbf{5.77} & 60.52 & 14.80 & 46.35 & \textbf{8.37} & 2.00 \\
\hline
\multirow{3}{*}{Actor} & Rand. & 23.84 & 1.15 & 24.04 & \textbf{0.84} & 30.95 & 1.15 & 24.40 & \textbf{0.61} & 33.33 & 1.59 & 31.37 & 1.41 & 33.72 & 1.15 & 2.14 \\
 & Class. & 23.91 & 0.96 & 24.41 & 1.19 & 30.56 & \textbf{1.07} & 24.82 & 1.21 & 33.04 & \textbf{0.60} & 30.52 & 0.98 & 33.23 & \textbf{1.02} & \textbf{1.86} \\
 & \hp{} & 24.58 & \textbf{0.46} & 24.48 & 0.96 & 30.74 & 2.00 & 24.80 & 1.05 & 32.91 & 0.68 & 30.72 & \textbf{0.83} & 33.31 & 1.42 & 2.00 \\
\hline
\multirow{3}{*}{Cornell} & Rand. & 24.70 & \textbf{3.40} & 24.77 & \textbf{3.45} & 54.64 & 11.57 & 21.29 & 2.45 & 38.81 & 10.63 & 59.26 & 9.75 & 28.99 & 14.37 & 1.86 \\
 & Class. & 27.44 & 5.98 & 23.18 & 5.85 & 48.76 & \textbf{7.98} & 21.01 & \textbf{1.66} & 39.72 & 9.87 & 57.25 & 12.83 & 21.56 & \textbf{1.89} & \textbf{1.71} \\
 & \hp{} & 24.05 & 7.11 & 27.98 & 7.34 & 52.63 & 12.02 & 21.50 & 3.36 & 43.38 & \textbf{8.97} & 56.10 & \textbf{9.69} & 30.69 & 16.65 & 2.43 \\
\hline
\multirow{3}{*}{Texas} & Rand. & 27.78 & \textbf{4.64} & 26.31 & 4.14 & 67.23 & 10.43 & 26.76 & \textbf{3.24} & 42.17 & 10.62 & 68.04 & 12.53 & 36.25 & 20.32 & 1.86 \\
 & Class. & 31.38 & 6.00 & 26.85 & 5.25 & 66.42 & 18.05 & 24.02 & 3.49 & 41.57 & \textbf{8.15} & 69.37 & 12.80 & 55.28 & 31.84 & 2.43 \\
 & \hp{} & 31.69 & 8.81 & 25.70 & \textbf{2.60} & 57.38 & \textbf{4.82} & 26.75 & 4.43 & 41.79 & 9.35 & 71.40 & \textbf{8.47} & 26.75 & \textbf{4.43} & \textbf{1.71} \\
\hline
\multirow{3}{*}{Roman} & Rand. & 38.17 & 0.57 & 28.96 & 0.76 & 64.47 & 0.82 & 37.74 & \textbf{0.20} & 63.81 & 0.62 & 71.91 & \textbf{0.78} & 58.31 & 0.55 & 2.14 \\
 & Class. & 38.25 & 0.53 & 29.33 & \textbf{0.74} & 64.42 & 0.78 & 37.54 & 0.80 & 63.50 & 0.72 & 71.37 & 0.89 & 58.14 & \textbf{0.42} & 2.14 \\
 & \hp{} & 38.47 & \textbf{0.42} & 29.32 & 0.76 & 64.39 & \textbf{0.40} & 37.55 & 0.44 & 63.81 & \textbf{0.59} & 71.46 & 0.82 & 58.50 & 0.66 & \textbf{1.71} \\
\midrule
\multirow{3}{*}{Avg.\ Rank} & Rand. & \multicolumn{2}{c|}{1.93} & \multicolumn{2}{c|}{2.00} & \multicolumn{2}{c|}{2.13} & \multicolumn{2}{c|}{2.13} & \multicolumn{2}{c|}{2.27} & \multicolumn{2}{c|}{2.07} & \multicolumn{2}{c|}{2.27} & 2.11 \\
 & Class. & \multicolumn{2}{c|}{2.47} & \multicolumn{2}{c|}{2.33} & \multicolumn{2}{c|}{2.07} & \multicolumn{2}{c|}{2.07} & \multicolumn{2}{c|}{2.33} & \multicolumn{2}{c|}{2.33} & \multicolumn{2}{c|}{2.20} & 2.26 \\
 & \hp{} & \multicolumn{2}{c|}{\textbf{1.60}} & \multicolumn{2}{c|}{\textbf{1.67}} & \multicolumn{2}{c|}{\textbf{1.80}} & \multicolumn{2}{c|}{\textbf{1.80}} & \multicolumn{2}{c|}{\textbf{1.40}} & \multicolumn{2}{c|}{\textbf{1.60}} & \multicolumn{2}{c|}{\textbf{1.53}} & \textbf{1.63} \\
\bottomrule
\end{tabular}}
\end{table}

\newpage

\begin{table}[h]
\centering
\caption{Mean test accuracy (\%) and cross-fold standard deviation (\(\times 10^2\)) at \(k=4\), averaged over 6 random seeds, for all datasets, models, and split strategies. \textbf{Bold} std = lowest per (dataset, model). Ratings = Amazon-Ratings; Roman = Roman-Empire.}
\label{tab:acc_per_dataset_k4_seeds}
\renewcommand{\arraystretch}{0.85}
\scriptsize
\resizebox{\textwidth}{!}{%
\begin{tabular}{ll|cc|cc|cc|cc|cc|cc|cc|c}
\toprule
 & & \multicolumn{2}{c|}{GCN} & \multicolumn{2}{c|}{GAT} & \multicolumn{2}{c|}{GraphSAGE} & \multicolumn{2}{c|}{APPNP} & \multicolumn{2}{c|}{MixHop} & \multicolumn{2}{c|}{H2GCN} & \multicolumn{2}{c|}{GPR-GNN} & \\
Dataset & Strategy & Mean & Std & Mean & Std & Mean & Std & Mean & Std & Mean & Std & Mean & Std & Mean & Std & Rank \\
\midrule
\multirow{3}{*}{Physics} & Rand. & 95.10 & 0.41 & 94.61 & 0.41 & 95.44 & 0.32 & 94.97 & 0.35 & 95.39 & \textbf{0.28} & 95.69 & \textbf{0.27} & 95.70 & 0.32 & 2.43 \\
 & Class. & 95.03 & 0.31 & 94.63 & 0.29 & 95.43 & \textbf{0.28} & 94.87 & 0.36 & 95.33 & 0.29 & 95.62 & 0.34 & 95.73 & \textbf{0.27} & 2.29 \\
 & \hp{} & 95.06 & \textbf{0.23} & 94.59 & \textbf{0.25} & 95.32 & 0.29 & 94.91 & \textbf{0.27} & 95.38 & \textbf{0.28} & 95.64 & 0.29 & 95.73 & \textbf{0.27} & \textbf{1.29} \\
\hline
\multirow{3}{*}{Photo} & Rand. & 92.61 & 0.80 & 93.34 & 0.67 & 92.82 & 1.04 & 93.36 & 0.70 & 94.04 & 0.64 & 93.40 & 0.78 & 94.18 & 0.57 & 2.29 \\
 & Class. & 92.61 & 0.78 & 93.38 & 0.85 & 92.99 & 0.93 & 93.33 & 0.76 & 93.89 & 0.68 & 93.35 & 0.81 & 94.08 & 0.61 & 2.71 \\
 & \hp{} & 92.81 & \textbf{0.41} & 93.49 & \textbf{0.56} & 93.12 & \textbf{0.52} & 93.51 & \textbf{0.57} & 94.11 & \textbf{0.58} & 93.50 & \textbf{0.56} & 94.25 & \textbf{0.42} & \textbf{1.00} \\
\hline
\multirow{3}{*}{CS} & Rand. & 91.33 & 0.65 & 90.66 & 0.64 & 91.85 & 0.63 & 91.71 & 0.69 & 91.95 & 0.72 & 93.13 & 0.67 & 92.76 & 0.60 & 2.43 \\
 & Class. & 91.46 & 0.63 & 90.72 & 0.75 & 91.82 & 0.66 & 91.69 & 0.69 & 92.01 & 0.58 & 93.35 & 0.55 & 92.90 & 0.67 & 2.57 \\
 & \hp{} & 91.35 & \textbf{0.48} & 90.56 & \textbf{0.49} & 91.77 & \textbf{0.34} & 91.72 & \textbf{0.48} & 92.00 & \textbf{0.45} & 93.26 & \textbf{0.46} & 92.77 & \textbf{0.45} & \textbf{1.00} \\
\hline
\multirow{3}{*}{Cora} & Rand. & 86.82 & 1.51 & 86.66 & 1.65 & 86.28 & 1.59 & 87.54 & 1.29 & 86.12 & 1.65 & 86.17 & 2.09 & 86.66 & 1.44 & 2.86 \\
 & Class. & 87.08 & 1.24 & 86.86 & 1.57 & 86.51 & 1.33 & 87.88 & 1.56 & 86.41 & 1.36 & 86.09 & 1.53 & 87.07 & \textbf{1.29} & 2.00 \\
 & \hp{} & 87.09 & \textbf{0.96} & 87.01 & \textbf{1.19} & 86.95 & \textbf{1.25} & 87.87 & \textbf{1.06} & 86.45 & \textbf{1.14} & 86.21 & \textbf{1.20} & 87.50 & 1.33 & \textbf{1.14} \\
\hline
\multirow{3}{*}{PubMed} & Rand. & 86.99 & \textbf{0.35} & 85.46 & 0.50 & 88.25 & 0.48 & 86.39 & 0.43 & 88.71 & 0.49 & 89.01 & 0.42 & 89.02 & 0.44 & 2.43 \\
 & Class. & 86.98 & 0.44 & 85.35 & 0.38 & 88.20 & 0.38 & 86.39 & 0.49 & 88.71 & 0.39 & 89.05 & 0.47 & 88.92 & 0.39 & 2.29 \\
 & \hp{} & 86.98 & 0.45 & 85.39 & \textbf{0.36} & 88.14 & \textbf{0.36} & 86.42 & \textbf{0.35} & 88.67 & \textbf{0.36} & 88.99 & \textbf{0.40} & 88.93 & \textbf{0.34} & \textbf{1.29} \\
\hline
\multirow{3}{*}{Computers} & Rand. & 88.71 & 0.90 & 90.27 & 0.80 & 85.53 & 4.65 & 88.12 & 1.12 & 90.32 & \textbf{1.01} & 89.88 & 1.21 & 90.07 & 0.88 & 2.29 \\
 & Class. & 88.77 & 0.93 & 90.44 & 0.67 & 86.28 & \textbf{4.29} & 87.97 & 0.92 & 90.41 & 1.13 & 90.39 & 0.84 & 90.09 & 1.08 & 2.14 \\
 & \hp{} & 88.64 & \textbf{0.63} & 90.35 & \textbf{0.54} & 84.88 & 4.84 & 87.89 & \textbf{0.74} & 90.74 & 1.31 & 90.35 & \textbf{0.66} & 90.14 & \textbf{0.81} & \textbf{1.57} \\
\hline
\multirow{3}{*}{CiteSeer} & Rand. & 72.21 & 2.00 & 71.61 & 1.64 & 70.92 & 1.85 & 71.91 & 1.49 & 70.19 & 1.34 & 69.69 & 1.70 & 71.22 & 1.95 & 2.86 \\
 & Class. & 72.29 & 1.38 & 71.41 & 1.25 & 71.11 & 1.53 & 72.02 & 1.41 & 70.19 & \textbf{1.25} & 69.68 & \textbf{1.39} & 71.05 & 1.34 & 1.71 \\
 & \hp{} & 72.19 & \textbf{1.36} & 71.66 & \textbf{1.18} & 71.11 & \textbf{1.12} & 71.83 & \textbf{1.12} & 70.56 & 1.35 & 69.23 & 1.61 & 71.14 & \textbf{1.27} & \textbf{1.43} \\
\hline
\multirow{3}{*}{Ratings} & Rand. & 26.97 & 0.40 & 25.50 & 0.39 & 29.38 & 0.53 & 26.29 & \textbf{0.25} & 33.42 & \textbf{0.87} & 31.09 & \textbf{1.01} & 30.65 & 0.64 & \textbf{1.71} \\
 & Class. & 26.98 & 0.53 & 25.56 & 0.56 & 29.55 & 0.51 & 26.31 & 0.39 & 33.63 & 1.19 & 31.23 & 1.05 & 30.75 & 0.80 & 2.71 \\
 & \hp{} & 27.00 & \textbf{0.36} & 25.37 & \textbf{0.22} & 29.51 & \textbf{0.35} & 26.31 & 0.35 & 33.27 & 1.05 & 31.23 & 1.18 & 31.07 & \textbf{0.39} & 1.57 \\
\hline
\multirow{3}{*}{Chameleon} & Rand. & 34.99 & 2.72 & 34.34 & 2.40 & 37.29 & \textbf{2.69} & 34.87 & 2.35 & 35.09 & \textbf{3.13} & 38.47 & 3.65 & 33.99 & 2.67 & 1.86 \\
 & Class. & 35.49 & \textbf{2.42} & 34.39 & \textbf{2.02} & 36.83 & 2.76 & 34.49 & \textbf{1.81} & 35.49 & 3.23 & 38.89 & 3.49 & 34.55 & \textbf{1.51} & \textbf{1.43} \\
 & \hp{} & 35.99 & 2.72 & 35.17 & 2.70 & 37.78 & 2.94 & 34.80 & 3.12 & 36.33 & 3.28 & 38.93 & \textbf{3.39} & 34.85 & 2.82 & 2.71 \\
\hline
\multirow{3}{*}{Squirrel} & Rand. & 26.19 & 1.95 & 23.88 & 1.46 & 28.12 & 1.63 & 26.79 & 1.52 & 28.59 & \textbf{2.07} & 29.95 & 1.63 & 26.37 & 1.60 & 2.29 \\
 & Class. & 25.19 & 1.80 & 23.26 & 1.37 & 27.54 & \textbf{1.50} & 25.82 & 1.41 & 27.86 & 2.16 & 29.41 & 2.10 & 25.89 & \textbf{1.54} & 2.00 \\
 & \hp{} & 25.43 & \textbf{1.42} & 23.49 & \textbf{1.13} & 27.52 & 2.12 & 26.73 & \textbf{1.29} & 27.52 & 2.11 & 29.19 & \textbf{1.51} & 26.92 & 1.63 & \textbf{1.71} \\
\hline
\multirow{3}{*}{Wisconsin} & Rand. & 27.67 & 5.14 & 29.15 & 7.03 & 56.17 & 11.10 & 27.16 & \textbf{3.60} & 41.57 & 10.91 & 62.36 & 9.60 & 52.50 & 17.82 & 2.71 \\
 & Class. & 28.46 & 4.65 & 31.62 & 6.65 & 55.42 & 8.00 & 27.80 & 3.77 & 41.72 & \textbf{7.25} & 61.90 & 7.84 & 47.94 & 15.87 & \textbf{1.86} \\
 & \hp{} & 27.76 & \textbf{3.91} & 30.04 & \textbf{5.19} & 57.68 & \textbf{7.74} & 27.89 & 3.78 & 42.24 & 7.36 & 60.37 & \textbf{5.71} & 46.05 & \textbf{15.14} & 1.43 \\
\hline
\multirow{3}{*}{Actor} & Rand. & 23.80 & 0.88 & 24.22 & 0.94 & 30.17 & \textbf{0.84} & 24.12 & 0.96 & 32.01 & 1.06 & 30.54 & 1.29 & 32.71 & 0.96 & 2.14 \\
 & Class. & 23.94 & \textbf{0.80} & 23.99 & 0.89 & 30.20 & 0.96 & 24.39 & 0.98 & 32.02 & 1.22 & 30.14 & 1.50 & 32.90 & 1.10 & 2.43 \\
 & \hp{} & 23.86 & \textbf{0.80} & 24.08 & \textbf{0.81} & 30.06 & 1.09 & 24.34 & \textbf{0.90} & 32.01 & \textbf{0.96} & 30.20 & \textbf{1.18} & 32.17 & \textbf{0.93} & \textbf{1.43} \\
\hline
\multirow{3}{*}{Cornell} & Rand. & 24.51 & \textbf{5.31} & 23.09 & \textbf{4.61} & 50.76 & 9.91 & 20.89 & 2.62 & 34.48 & 10.81 & 57.72 & 8.05 & 25.13 & 11.03 & 2.00 \\
 & Class. & 24.53 & 5.77 & 25.38 & 7.72 & 50.60 & \textbf{7.21} & 21.42 & \textbf{2.57} & 40.29 & \textbf{10.54} & 56.20 & \textbf{6.58} & 24.89 & 9.35 & \textbf{1.71} \\
 & \hp{} & 23.87 & 5.39 & 24.99 & 6.06 & 55.41 & 12.01 & 21.49 & 3.13 & 34.35 & 12.70 & 59.21 & 7.11 & 25.08 & \textbf{8.92} & 2.29 \\
\hline
\multirow{3}{*}{Texas} & Rand. & 26.73 & 6.81 & 27.11 & 7.41 & 60.00 & 14.48 & 25.45 & 3.70 & 35.80 & \textbf{6.81} & 66.99 & \textbf{8.81} & 30.88 & \textbf{13.93} & 2.00 \\
 & Class. & 28.11 & 6.08 & 28.64 & 5.08 & 59.49 & \textbf{10.16} & 27.50 & 4.98 & 34.46 & 7.64 & 68.28 & 11.82 & 32.64 & 15.08 & 2.14 \\
 & \hp{} & 26.56 & \textbf{4.58} & 26.63 & \textbf{4.13} & 58.88 & 11.44 & 24.97 & \textbf{2.57} & 40.14 & 10.90 & 69.08 & 10.23 & 33.88 & 16.15 & \textbf{1.86} \\
\hline
\multirow{3}{*}{Roman} & Rand. & 38.26 & 0.73 & 29.13 & 0.93 & 64.49 & 0.69 & 37.68 & 0.85 & 63.40 & 1.09 & 71.26 & 0.67 & 58.30 & 0.79 & 2.86 \\
 & Class. & 38.11 & 0.68 & 29.13 & \textbf{0.68} & 64.38 & 0.59 & 37.68 & 0.75 & 63.13 & 0.88 & 71.56 & 0.67 & 58.42 & \textbf{0.52} & 1.86 \\
 & \hp{} & 38.26 & \textbf{0.61} & 29.10 & 0.91 & 64.56 & \textbf{0.55} & 37.71 & \textbf{0.57} & 63.35 & \textbf{0.75} & 71.44 & \textbf{0.59} & 58.44 & 0.62 & \textbf{1.29} \\
\midrule
\multirow{3}{*}{Avg.\ Rank} & Rand. & \multicolumn{2}{c|}{2.53} & \multicolumn{2}{c|}{2.60} & \multicolumn{2}{c|}{2.47} & \multicolumn{2}{c|}{2.13} & \multicolumn{2}{c|}{2.00} & \multicolumn{2}{c|}{2.27} & \multicolumn{2}{c|}{2.40} & 2.34 \\
 & Class. & \multicolumn{2}{c|}{2.07} & \multicolumn{2}{c|}{2.13} & \multicolumn{2}{c|}{1.73} & \multicolumn{2}{c|}{2.40} & \multicolumn{2}{c|}{2.13} & \multicolumn{2}{c|}{2.33} & \multicolumn{2}{c|}{2.07} & 2.12 \\
 & \hp{} & \multicolumn{2}{c|}{\textbf{1.40}} & \multicolumn{2}{c|}{\textbf{1.27}} & \multicolumn{2}{c|}{\textbf{1.80}} & \multicolumn{2}{c|}{\textbf{1.47}} & \multicolumn{2}{c|}{\textbf{1.87}} & \multicolumn{2}{c|}{\textbf{1.40}} & \multicolumn{2}{c|}{\textbf{1.53}} & \textbf{1.53} \\
\bottomrule
\end{tabular}}
\end{table}

\newpage

\begin{table}[h]
\centering
\caption{Average reduction in cross-fold standard deviation of test accuracy relative to random \(k\)-fold, averaged across all 7 models per dataset, for class-stratified \(k\)-fold and \hp{} under \(k=4\), \(k=5\), and \(k=4\) averaged over 6 seeds. Green indicates a positive delta, where the method achieves lower variance than random \(k\)-fold; this holds for the majority of dataset/method combinations. The final row reports the average across all 15 datasets.}
\label{tab:stddevdelta}
\renewcommand{\arraystretch}{0.85}
\begin{tabular}{l|l| ccc | ccc | ccc}
\toprule
 & & \multicolumn{3}{c|}{$k=4$} & \multicolumn{3}{c|}{$k=5$} & \multicolumn{3}{c}{$k=4$ (multi-seed)} \\
Dataset & Strategy & Rank & \multicolumn{2}{c|}{$\Delta$ (pp)} & Rank & \multicolumn{2}{c|}{$\Delta$ (pp)} & Rank & \multicolumn{2}{c}{$\Delta$ (pp)} \\
\midrule
\multirow{3}{*}{Physics} & Rand.   & 2.86 & \multicolumn{2}{c|}{---} & 2.14 & \multicolumn{2}{c|}{---} & 2.43 & \multicolumn{2}{c}{---} \\
 & Class.  & 1.71 & \multicolumn{2}{c|}{0.14} & 2.29 & \multicolumn{2}{c|}{-0.02} & 2.29 & \multicolumn{2}{c}{0.03} \\
 & HpStrat & \textbf{1.43} & \multicolumn{2}{c|}{\textcolor{green!60!black}{0.14}} & \textbf{1.57} & \multicolumn{2}{c|}{\textcolor{green!60!black}{0.03}} & \textbf{1.29} & \multicolumn{2}{c}{\textcolor{green!60!black}{0.07}} \\
\hline
\multirow{3}{*}{Photo} & Rand.   & 1.86 & \multicolumn{2}{c|}{---} & 2.14 & \multicolumn{2}{c|}{---} & 2.29 & \multicolumn{2}{c}{---} \\
 & Class.  & 3.00 & \multicolumn{2}{c|}{-0.39} & 2.86 & \multicolumn{2}{c|}{-0.40} & 2.71 & \multicolumn{2}{c}{-0.03} \\
 & HpStrat & \textbf{1.14} & \multicolumn{2}{c|}{\textcolor{green!60!black}{0.57}} & \textbf{1.00} & \multicolumn{2}{c|}{\textcolor{green!60!black}{0.45}} & \textbf{1.00} & \multicolumn{2}{c}{\textcolor{green!60!black}{0.22}} \\
\hline
\multirow{3}{*}{CS} & Rand.   & 2.14 & \multicolumn{2}{c|}{---} & 2.00 & \multicolumn{2}{c|}{---} & 2.43 & \multicolumn{2}{c}{---} \\
 & Class.  & 2.86 & \multicolumn{2}{c|}{-0.25} & 2.43 & \multicolumn{2}{c|}{-0.11} & 2.57 & \multicolumn{2}{c}{0.01} \\
 & HpStrat & \textbf{1.00} & \multicolumn{2}{c|}{\textcolor{green!60!black}{0.22}} & \textbf{1.57} & \multicolumn{2}{c|}{\textcolor{green!60!black}{0.05}} & \textbf{1.00} & \multicolumn{2}{c}{\textcolor{green!60!black}{0.21}} \\
\hline
\multirow{3}{*}{Cora} & Rand.   & 2.86 & \multicolumn{2}{c|}{---} & 2.14 & \multicolumn{2}{c|}{---} & 2.86 & \multicolumn{2}{c}{---} \\
 & Class.  & 2.00 & \multicolumn{2}{c|}{0.48} & 2.29 & \multicolumn{2}{c|}{-0.16} & 2.00 & \multicolumn{2}{c}{0.19} \\
 & HpStrat & \textbf{1.14} & \multicolumn{2}{c|}{\textcolor{green!60!black}{1.21}} & \textbf{1.57} & \multicolumn{2}{c|}{\textcolor{green!60!black}{0.12}} & \textbf{1.14} & \multicolumn{2}{c}{\textcolor{green!60!black}{0.44}} \\
\hline
\multirow{3}{*}{PubMed} & Rand.   & 2.71 & \multicolumn{2}{c|}{---} & 2.43 & \multicolumn{2}{c|}{---} & 2.43 & \multicolumn{2}{c}{---} \\
 & Class.  & 1.86 & \multicolumn{2}{c|}{0.29} & 2.57 & \multicolumn{2}{c|}{-0.01} & 2.29 & \multicolumn{2}{c}{0.03} \\
 & HpStrat & \textbf{1.43} & \multicolumn{2}{c|}{\textcolor{green!60!black}{0.34}} & \textbf{1.00} & \multicolumn{2}{c|}{\textcolor{green!60!black}{0.43}} & \textbf{1.29} & \multicolumn{2}{c}{\textcolor{green!60!black}{0.07}} \\
\hline
\multirow{3}{*}{Computers} & Rand.   & 2.29 & \multicolumn{2}{c|}{---} & 1.86 & \multicolumn{2}{c|}{---} & 2.29 & \multicolumn{2}{c}{---} \\
 & Class.  & 2.14 & \multicolumn{2}{c|}{-0.27} & 2.86 & \multicolumn{2}{c|}{-0.24} & 2.14 & \multicolumn{2}{c}{0.10} \\
 & HpStrat & \textbf{1.57} & \multicolumn{2}{c|}{-0.07} & \textbf{1.29} & \multicolumn{2}{c|}{\textcolor{green!60!black}{0.28}} & \textbf{1.57} & \multicolumn{2}{c}{\textcolor{green!60!black}{0.15}} \\
\hline
\multirow{3}{*}{CiteSeer} & Rand.   & 2.71 & \multicolumn{2}{c|}{---} & 2.71 & \multicolumn{2}{c|}{---} & 2.86 & \multicolumn{2}{c}{---} \\
 & Class.  & 1.71 & \multicolumn{2}{c|}{0.27} & 1.71 & \multicolumn{2}{c|}{0.58} & 1.71 & \multicolumn{2}{c}{0.34} \\
 & HpStrat & \textbf{1.57} & \multicolumn{2}{c|}{\textcolor{green!60!black}{0.51}} & \textbf{1.57} & \multicolumn{2}{c|}{\textcolor{green!60!black}{0.69}} & \textbf{1.43} & \multicolumn{2}{c}{\textcolor{green!60!black}{0.42}} \\
\hline
\multirow{3}{*}{Ratings} & Rand.   & 2.14 & \multicolumn{2}{c|}{---} & \textbf{1.71} & \multicolumn{2}{c|}{---} & 1.71 & \multicolumn{2}{c}{---} \\
 & Class.  & 2.29 & \multicolumn{2}{c|}{-0.07} & 2.43 & \multicolumn{2}{c|}{-0.11} & 2.43 & \multicolumn{2}{c}{-0.13} \\
 & HpStrat & \textbf{1.57} & \multicolumn{2}{c|}{\textcolor{green!60!black}{0.12}} & 1.86 & \multicolumn{2}{c|}{\textcolor{green!60!black}{0.02}} & \textbf{1.57} & \multicolumn{2}{c}{\textcolor{green!60!black}{0.03}} \\
\hline
\multirow{3}{*}{Chameleon} & Rand.   & 2.14 & \multicolumn{2}{c|}{---} & 2.43 & \multicolumn{2}{c|}{---} & 1.86 & \multicolumn{2}{c}{---} \\
 & Class.  & 2.57 & \multicolumn{2}{c|}{-0.11} & 2.00 & \multicolumn{2}{c|}{0.44} & \textbf{1.43} & \multicolumn{2}{c}{0.34} \\
 & HpStrat & \textbf{1.29} & \multicolumn{2}{c|}{\textcolor{green!60!black}{0.99}} & \textbf{1.57} & \multicolumn{2}{c|}{\textcolor{green!60!black}{0.49}} & 2.71 & \multicolumn{2}{c}{-0.20} \\
\hline
\multirow{3}{*}{Squirrel} & Rand.   & 2.71 & \multicolumn{2}{c|}{---} & 2.29 & \multicolumn{2}{c|}{---} & 2.29 & \multicolumn{2}{c}{---} \\
 & Class.  & 1.86 & \multicolumn{2}{c|}{0.32} & 2.14 & \multicolumn{2}{c|}{0.35} & 2.00 & \multicolumn{2}{c}{-0.00} \\
 & HpStrat & \textbf{1.43} & \multicolumn{2}{c|}{\textcolor{green!60!black}{0.36}} & \textbf{1.57} & \multicolumn{2}{c|}{\textcolor{green!60!black}{0.80}} & \textbf{1.71} & \multicolumn{2}{c}{\textcolor{green!60!black}{0.09}} \\
\hline
\multirow{3}{*}{Wisconsin} & Rand.   & \textbf{1.86} & \multicolumn{2}{c|}{---} & \textbf{1.86} & \multicolumn{2}{c|}{---} & 2.71 & \multicolumn{2}{c}{---} \\
 & Class.  & \textbf{1.86} & \multicolumn{2}{c|}{0.35} & 2.14 & \multicolumn{2}{c|}{-0.60} & 1.86 & \multicolumn{2}{c}{1.60} \\
 & HpStrat & 2.29 & \multicolumn{2}{c|}{-0.21} & 2.00 & \multicolumn{2}{c|}{-0.07} & \textbf{1.43} & \multicolumn{2}{c}{\textcolor{green!60!black}{2.34}} \\
\hline
\multirow{3}{*}{Actor} & Rand.   & 2.14 & \multicolumn{2}{c|}{---} & 2.14 & \multicolumn{2}{c|}{---} & 2.14 & \multicolumn{2}{c}{---} \\
 & Class.  & \textbf{1.86} & \multicolumn{2}{c|}{0.07} & \textbf{1.86} & \multicolumn{2}{c|}{0.12} & 2.43 & \multicolumn{2}{c}{-0.07} \\
 & HpStrat & 2.00 & \multicolumn{2}{c|}{0.01} & 2.00 & \multicolumn{2}{c|}{0.07} & \textbf{1.43} & \multicolumn{2}{c}{\textcolor{green!60!black}{0.04}} \\
\hline
\multirow{3}{*}{Cornell} & Rand.   & 2.57 & \multicolumn{2}{c|}{---} & 1.86 & \multicolumn{2}{c|}{---} & 2.00 & \multicolumn{2}{c}{---} \\
 & Class.  & 2.00 & \multicolumn{2}{c|}{1.76} & \textbf{1.71} & \multicolumn{2}{c|}{1.37} & \textbf{1.71} & \multicolumn{2}{c}{0.37} \\
 & HpStrat & \textbf{1.43} & \multicolumn{2}{c|}{\textcolor{green!60!black}{3.38}} & 2.43 & \multicolumn{2}{c|}{-1.36} & 2.29 & \multicolumn{2}{c}{-0.42} \\
\hline
\multirow{3}{*}{Texas} & Rand.   & 1.86 & \multicolumn{2}{c|}{---} & 1.86 & \multicolumn{2}{c|}{---} & 2.00 & \multicolumn{2}{c}{---} \\
 & Class.  & 2.71 & \multicolumn{2}{c|}{-0.50} & 2.43 & \multicolumn{2}{c|}{-2.81} & 2.14 & \multicolumn{2}{c}{0.16} \\
 & HpStrat & \textbf{1.43} & \multicolumn{2}{c|}{\textcolor{green!60!black}{0.37}} & \textbf{1.71} & \multicolumn{2}{c|}{\textcolor{green!60!black}{3.29}} & \textbf{1.86} & \multicolumn{2}{c}{\textcolor{green!60!black}{0.28}} \\
\hline
\multirow{3}{*}{Roman} & Rand.   & 1.86 & \multicolumn{2}{c|}{---} & 2.14 & \multicolumn{2}{c|}{---} & 2.86 & \multicolumn{2}{c}{---} \\
 & Class.  & 2.57 & \multicolumn{2}{c|}{-0.26} & 2.14 & \multicolumn{2}{c|}{-0.08} & 1.86 & \multicolumn{2}{c}{0.14} \\
 & HpStrat & \textbf{1.57} & \multicolumn{2}{c|}{0.00} & \textbf{1.71} & \multicolumn{2}{c|}{\textcolor{green!60!black}{0.03}} & \textbf{1.29} & \multicolumn{2}{c}{\textcolor{green!60!black}{0.17}} \\
\midrule
\multirow{3}{*}{\textbf{Average}} & Rand.   & 2.31 & \multicolumn{2}{c|}{---} & 2.11 & \multicolumn{2}{c|}{---} & 2.34 & \multicolumn{2}{c}{---} \\
 & Class.  & 2.20 & \multicolumn{2}{c|}{0.12} & 2.26 & \multicolumn{2}{c|}{-0.11} & 2.12 & \multicolumn{2}{c}{0.20} \\
 & HpStrat & \textbf{1.49} & \multicolumn{2}{c|}{\textcolor{green!60!black}{0.53}} & \textbf{1.63} & \multicolumn{2}{c|}{\textcolor{green!60!black}{0.36}} & \textbf{1.53} & \multicolumn{2}{c}{\textcolor{green!60!black}{0.26}} \\
\bottomrule
\end{tabular}
\end{table}

\newpage
\section{Per-Dataset Fold Quality Results}
\label{app:foldmetrics_per_dataset}

Table~\ref{tab:foldmetrics} reports the three fold-quality metrics for all 15 datasets and 3 split strategies, with each value reported as mean $\pm$ standard deviation across 10 seeds. All metrics are lower-is-better. Values are scaled for readability: SDD is shown as-is (\(\times 10^{0}\)), CDD is multiplied by \(10^{3}\), CHD by \(10^{4}\). Bold values indicate the lowest mean value per (dataset, metric) across strategies. The Rank column reports the mean rank of each strategy across all three metrics for that dataset; the Avg.\ Rank row reports mean ranks across all 15 datasets.

\begin{table}[h]
\centering
\caption{Fold-quality metrics for all datasets and split strategies. Lower is better. SDD \(\times 10^{0}\); CDD \(\times 10^{0}\); CHD \(\times 10^{2}\). \textbf{Bold} = lowest mean per (dataset, metric). Ratings = Amazon-Ratings; Roman = Roman-Empire.}
\label{tab:foldmetrics}
\renewcommand{\arraystretch}{0.85}
\setlength{\tabcolsep}{3pt}
\scriptsize
\begin{tabular}{ll|cc|cc|cc}
\toprule
 & & \multicolumn{2}{c|}{SDD} & \multicolumn{2}{c|}{CDD} & \multicolumn{2}{c}{CHD} \\
\cmidrule(lr){3-4}\cmidrule(lr){5-6}\cmidrule(lr){7-8}
Dataset & Strategy & Mean & Std & Mean & Std & Mean & Std \\
\midrule
\multirow{3}{*}{Physics} & Rand.   & \textbf{0.38} & 0.00 & 21.48 & 3.16 & 0.13 & 0.06 \\
 & Class.                          & \textbf{0.38} & 0.00 & \textbf{0.42} & 0.00 & 0.12 & 0.06 \\
 & \hp{}                           & \textbf{0.38} & 0.00 & 1.02  & 0.00 & \textbf{0.02} & 0.01 \\
\hline
\multirow{3}{*}{Photo} & Rand.     & \textbf{0.50} & 0.00 & 9.09  & 1.27 & 0.32 & 0.15 \\
 & Class.                          & \textbf{0.50} & 0.00 & \textbf{0.41} & 0.00 & 0.49 & 0.15 \\
 & \hp{}                           & \textbf{0.50} & 0.00 & 0.56  & 0.00 & \textbf{0.04} & 0.02 \\
\hline
\multirow{3}{*}{CS} & Rand.        & \textbf{0.38} & 0.00 & 10.58 & 1.35 & 0.16 & 0.07 \\
 & Class.                          & \textbf{0.38} & 0.00 & \textbf{0.26} & 0.00 & 0.23 & 0.09 \\
 & \hp{}                           & \textbf{0.38} & 0.00 & 0.97  & 0.00 & \textbf{0.03} & 0.01 \\
\hline
\multirow{3}{*}{Cora} & Rand.      & \textbf{0.00} & 0.00 & 6.10  & 0.76 & 1.07 & 0.31 \\
 & Class.                          & \textbf{0.00} & 0.00 & \textbf{0.39} & 0.00 & 0.83 & 0.30 \\
 & \hp{}                           & \textbf{0.00} & 0.00 & 1.43  & 0.00 & \textbf{0.05} & 0.02 \\
\hline
\multirow{3}{*}{PubMed} & Rand.    & \textbf{0.38} & 0.00 & 19.82 & 8.00 & 0.37 & 0.12 \\
 & Class.                          & \textbf{0.38} & 0.00 & \textbf{0.38} & 0.00 & 0.33 & 0.12 \\
 & \hp{}                           & \textbf{0.38} & 0.00 & 1.12  & 0.00 & \textbf{0.02} & 0.01 \\
\hline
\multirow{3}{*}{Computers} & Rand. & \textbf{0.00} & 0.00 & 9.67  & 1.42 & 0.26 & 0.12 \\
 & Class.                          & \textbf{0.00} & 0.00 & \textbf{0.33} & 0.00 & 0.28 & 0.12 \\
 & \hp{}                           & \textbf{0.00} & 0.00 & 0.82  & 0.00 & \textbf{0.04} & 0.01 \\
\hline
\multirow{3}{*}{CiteSeer} & Rand.  & \textbf{0.38} & 0.00 & 7.16  & 1.40 & 1.01 & 0.43 \\
 & Class.                          & \textbf{0.38} & 0.00 & \textbf{0.15} & 0.00 & 0.70 & 0.24 \\
 & \hp{}                           & \textbf{0.38} & 0.00 & 1.27  & 0.00 & \textbf{0.04} & 0.01 \\
\hline
\multirow{3}{*}{Ratings} & Rand.   & \textbf{0.00} & 0.00 & 18.58 & 5.14 & 0.27 & 0.09 \\
 & Class.                          & \textbf{0.00} & 0.00 & \textbf{0.35} & 0.00 & 0.26 & 0.10 \\
 & \hp{}                           & \textbf{0.00} & 0.00 & 1.10  & 0.00 & \textbf{0.02} & 0.01 \\
\hline
\multirow{3}{*}{Chameleon} & Rand. & \textbf{0.50} & 0.00 & 3.69  & 0.79 & 0.98 & 0.77 \\
 & Class.                          & \textbf{0.50} & 0.00 & \textbf{0.35} & 0.00 & 1.05 & 0.49 \\
 & \hp{}                           & \textbf{0.50} & 0.00 & 1.02  & 0.00 & \textbf{0.11} & 0.05 \\
\hline
\multirow{3}{*}{Squirrel} & Rand.  & \textbf{0.38} & 0.00 & 5.96  & 1.52 & 0.51 & 0.24 \\
 & Class.                          & \textbf{0.38} & 0.00 & \textbf{0.23} & 0.00 & 0.40 & 0.19 \\
 & \hp{}                           & \textbf{0.38} & 0.00 & 1.07  & 0.00 & \textbf{0.08} & 0.04 \\
\hline
\multirow{3}{*}{Wisconsin} & Rand. & \textbf{0.38} & 0.00 & 1.82  & 0.35 & 2.37 & 0.98 \\
 & Class.                          & \textbf{0.38} & 0.00 & \textbf{0.38} & 0.00 & 1.93 & 0.60 \\
 & \hp{}                           & \textbf{0.38} & 0.00 & 0.90  & 0.00 & \textbf{0.40} & 0.08 \\
\hline
\multirow{3}{*}{Actor} & Rand.     & \textbf{0.00} & 0.00 & 12.77 & 3.77 & 0.40 & 0.23 \\
 & Class.                          & \textbf{0.00} & 0.00 & \textbf{0.40} & 0.00 & 0.43 & 0.20 \\
 & \hp{}                           & \textbf{0.00} & 0.00 & 0.65  & 0.00 & \textbf{0.03} & 0.01 \\
\hline
\multirow{3}{*}{Cornell} & Rand.   & \textbf{0.38} & 0.00 & 1.58  & 0.18 & 3.22 & 0.99 \\
 & Class.                          & \textbf{0.38} & 0.00 & \textbf{0.38} & 0.00 & 2.41 & 0.79 \\
 & \hp{}                           & \textbf{0.38} & 0.00 & 0.55  & 0.00 & \textbf{0.31} & 0.05 \\
\hline
\multirow{3}{*}{Texas} & Rand.     & \textbf{0.38} & 0.00 & 1.38  & 0.39 & 1.58 & 0.63 \\
 & Class.                          & \textbf{0.38} & 0.00 & \textbf{0.42} & 0.00 & 1.84 & 0.69 \\
 & \hp{}                           & \textbf{0.38} & 0.00 & 0.78  & 0.00 & \textbf{0.63} & 0.00 \\
\hline
\multirow{3}{*}{Roman} & Rand.     & \textbf{0.50} & 0.00 & 10.88 & 0.90 & 0.11 & 0.03 \\
 & Class.                          & \textbf{0.50} & 0.00 & \textbf{0.26} & 0.00 & 0.13 & 0.05 \\
 & \hp{}                           & \textbf{0.50} & 0.00 & 0.56  & 0.00 & \textbf{0.01} & 0.00 \\
\midrule
\multirow{3}{*}{Avg.\ Rank} & Rand.  & \multicolumn{2}{c|}{\textbf{1.00}} & \multicolumn{2}{c|}{3.00} & \multicolumn{2}{c}{2.53} \\
 & Class.                            & \multicolumn{2}{c|}{\textbf{1.00}} & \multicolumn{2}{c|}{\textbf{1.00}} & \multicolumn{2}{c}{2.47} \\
 & \hp{}                             & \multicolumn{2}{c|}{\textbf{1.00}} & \multicolumn{2}{c|}{\textit{2.00}} & \multicolumn{2}{c}{\textbf{1.00}} \\\bottomrule
\end{tabular}
\end{table}
\section{Effect of Split Strategy on Model Selection}
\label{app:best_model_shift}

We motivate \hp{} by its effect on cross-fold variance, but reduced variance also has a direct practical consequence: it affects which architecture is selected as best on a given dataset. A practitioner comparing architectures on a new graph dataset typically evaluates them under one of a small number of standard protocols, such as averaging over one or several random seeds~\cite{pitfallspaper}, random \(k\)-fold, or class-stratified \(k\)-fold. Prior work~\cite{pitfallspaper} shows that the identity of the best-performing model already shifts depending on the random seed used to draw a single split. Table~\ref{tab:best_model_shift} shows that this instability extends across split strategies more broadly: for each of the 15 datasets, we report the best-performing architecture, by mean test accuracy, under random \(k\)-fold, class-stratified \(k\)-fold, and \hp{}.

The best-performing architecture differs across at least two of the three strategies on 8 of the 15 datasets. When cross-fold standard deviation is high, the ranking of architectures by mean accuracy becomes sensitive to the split used to compute that mean, so different splitting strategies can produce different, equally plausible, rankings. We showed in Section~\ref{sec:results:accuracy} that \hp{} reduces this variance. The result here shows the practical consequence of that reduction: lower variance changes which architecture a benchmark reports as best.

\begin{table}[t]
\centering
\caption{Best-performing architecture, by mean test accuracy, under random \(k\)-fold, class-stratified \(k\)-fold, and \hp{}, for all 15 datasets. The best model differs across at least two of the three strategies on 8 of 15 datasets.}
\label{tab:best_model_shift}
\begin{tabular}{lcccc}
\toprule
Dataset & Random & Class-stratified & \hp{} & Consistent? \\
\midrule
Physics        & H2GCN     & GPR-GNN & GPR-GNN   & No \\
Photo          & MixHop    & GPR-GNN & GPR-GNN   & No \\
CS             & H2GCN     & H2GCN   & H2GCN     & Yes \\
Cora           & APPNP     & APPNP   & APPNP     & Yes \\
PubMed         & MixHop    & H2GCN   & GPR-GNN   & No \\
Computers      & MixHop    & H2GCN   & MixHop    & No \\
CiteSeer       & GPR-GNN   & GCN     & GCN       & No \\
Amazon-Ratings & MixHop    & MixHop  & MixHop    & Yes \\
Chameleon      & MixHop    & MixHop  & H2GCN     & No \\
Squirrel       & H2GCN     & H2GCN   & H2GCN     & Yes \\
Wisconsin      & H2GCN     & H2GCN   & H2GCN     & Yes \\
Actor          & GPR-GNN   & GPR-GNN & GPR-GNN   & Yes \\
Cornell        & H2GCN     & H2GCN   & GraphSAGE & No \\
Texas          & GraphSAGE & H2GCN   & H2GCN     & No \\
Roman-Empire   & H2GCN     & H2GCN   & H2GCN     & Yes \\
\bottomrule
\end{tabular}
\end{table}   
\section{Runtime Analysis}
\label{app:runtime}

\begin{figure}[h]
\centering
\includegraphics[width=0.45\textwidth]{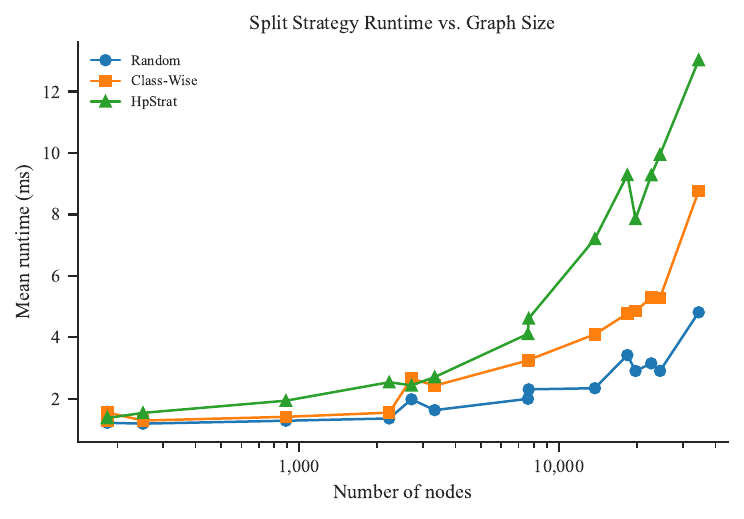}
\caption{Mean split construction runtime (ms) versus graph size for all three splitting strategies across the 15 benchmark datasets, measured on an Apple MacBook M3 Pro with 36\,GB of memory. All methods remain in the single-digit millisecond range. The additional cost of \hp{} relative to the baselines is negligible compared to the cost of model training.}
\label{fig:runtime}
\end{figure}

\hp{} introduces a homophily computation step that random $k$-fold and class-stratified $k$-fold do not require. Figure~\ref{fig:runtime} reports mean split construction time across all 15 datasets as a function of graph size, measured on an Apple MacBook M3 Pro with 36\,GB of memory. All three methods operate in the single-digit millisecond range across the full range of graph sizes evaluated, from the smallest datasets (fewer than 200 nodes) to the largest (approximately 35{,}000 nodes). \hp{} is slower than both baselines at larger graph sizes, with the gap widening with node count, but remains below 13\,ms even for the largest graph evaluated. Since split construction is a one-time preprocessing step executed once per graph before any model training, this overhead is negligible in practice relative to the cost of model training runs.

\newpage
\section{Sensitivity to Bin Count}
\label{app:bin_ablation}

\hp{} introduces one hyperparameter, the number of homophily bins $B$. To verify that the headline result is not sensitive to this choice, we ran \hp{} on GCN/Cora with $B \in \{5, 10, 20, 50\}$, alongside random $k$-fold and class-stratified $k$-fold as baselines. Cross-fold standard deviation is reported in Table~\ref{tab:bin_ablation}.

\begin{table}[h]
\centering
\caption{Sensitivity of \hp{} to bin count $B$ on GCN/Cora. Cross-fold standard deviation is reported as $\times 100$ for readability. \hp{} beats random $k$-fold for every value of $B$ tested, and beats class-stratified $k$-fold for $B \leq 20$. Bold = lowest std.}
\label{tab:bin_ablation}
\begin{tabular}{lrrr}
\toprule
Strategy & $B$ & Mean Acc (\%) & Std ($\times 100$) \\
\midrule
Random           & ---  & 85.94 & 2.79 \\
Class-stratified & ---  & 86.83 & 1.75 \\
\hp{}            & 5    & 86.80 & 1.15 \\
\hp{}            & 10   & 86.10 & \textbf{0.90} \\
\hp{}            & 20   & 86.47 & 0.97 \\
\hp{}            & 50   & 86.83 & 2.05 \\
\bottomrule
\end{tabular}
\end{table}

\hp{} produces lower cross-fold standard deviation than random $k$-fold across all four values of $B$, confirming that the headline result is insensitive to the choice of $B$ within a sensible range. Against class-stratified $k$-fold, \hp{} wins for $B \in \{5, 10, 20\}$ but degrades at $B = 50$. This is the small-stratum regime: with $|\mathcal{V}| = 2{,}708$ nodes, $C = 7$ classes, $k = 4$ folds, and $B = 50$ bins, the average bin-class stratum holds $|\mathcal{V}| / (B \cdot C \cdot k) \approx 2$ nodes, leaving most strata too sparse for round-robin assignment to achieve proportional representation. The procedure does not fail in this regime; it converges toward less informative assignments because there is little structure left to exploit within each stratum. The main experiments use $B = 10$, which lies safely outside this regime for all 15 datasets evaluated.

\end{document}